\documentclass[fleqn,usenatbib]{mnras}

\usepackage{newtxtext,newtxmath}
\usepackage[T1]{fontenc}
\usepackage{ae,aecompl}

\usepackage{graphicx}	% Including figure files
\usepackage{amsmath}	% Advanced maths commands
\usepackage{amssymb}	% Extra maths symbols
\usepackage{adjustbox}
\usepackage[flushleft]{threeparttable}
\usepackage{booktabs,caption}
\usepackage[normalem]{ulem}

\newcommand{\bh}{{\rm BH}}
\newcommand{\bhs}{{\rm BHS}}

\newcommand{\Ms}{~{\rm M}_\odot}

\newcommand{\Log}{{\rm Log}}

\title[Galactic GCs Harbouring a BH Subsystem]{MOCCA-SURVEY Database I: Galactic Globular Clusters Harbouring a Black Hole Subsystem}

\author[A.Askar, M. Arca Sedda, M. Giersz]{
Abbas Askar$^{1}$\thanks{E-mail: askar@camk.edu.pl},
Manuel Arca Sedda$^{2}$
and Mirek Giersz$^{1}$
\\
$^{1}$Nicolaus Copernicus Astronomical Center, Polish Academy of Sciences, 
ul. Bartycka 18, 00-716 Warsaw, Poland\\
$^{2}$Astronomisches Rechen-Institut, Zentrum f{\"u}r Astronomie, University of Heidelberg, M{\"o}nchhofstrasse 12-14, 69120, Heidelberg, Germany\\
}

\date{Accepted XXX. Received YYY; in original form ZZZ}

\pubyear{2015}

\begin{document}
\label{firstpage}
\pagerange{\pageref{firstpage}--\pageref{lastpage}}
\maketitle

\begin{abstract}
There have been increasing theoretical speculations and observational indications that certain globular clusters (GCs) could contain a sizeable population of stellar mass black holes (BHs). In this paper, we shortlist at least 29 Galactic GCs that could be hosting a subsystem of BHs (BHS). In a companion paper, we analysed results from a wide array of GC models (simulated with the MOCCA code for cluster simulations) that retained few tens to several hundreds of BHs at 12 Gyr and showed that the properties of the BHS in those GCs correlate with the GC's observable properties. Building on those results, we use available observational properties of 140 Galactic GCs to identify 29 GCs that could potentially be harbouring up to a few hundreds of BHs. Utilizing observational properties and theoretical scaling relations, we estimate the density, size and mass of the BHS in these GCs. We also calculate the total number of BHs and the fraction of BHs contained in a binary system for our shortlisted Galactic GCs.
Additionally, we mention other Galactic GCs that could also contain significant number of single BHs or BHs in binary systems.
\end{abstract}


%%%%%%%%%%%%%%%%%%%%%%%%%%%%%%%%%%%%%%%%%%%%%%%%%%

%%%%%%%%%%%%%%%%% BODY OF PAPER %%%%%%%%%%%%%%%%%%

\section{Introduction}

Up to several thousands of black holes (BHs) may form in the most
massive globular clusters (GCs) from the evolution of massive stars
within a few Myr. Whether a large number of these BHs can be retained
in the GC depends significantly on the natal kicks that BHs
receive at birth and the escape velocity of the cluster. The exact magnitude or distribution of BH natal kicks is uncertain \citep{bkb2002,kb2010,repetto2012,fryer2012,janka13,mandel16,rep2017,rgw17} and initial properties of GCs are weakly constrained.  Whether GCs can sustain a sizeable population of BHs during their long term evolution over a dozen Gyr has also been deeply debated. Theoretical studies in early 1990s \citep{kulkarni93,sigurdsson93} had suggested that BHs that would be retained in GCs would quickly segregate to the center of the cluster forming a BH subsystem (BHS) that would decouple from the rest of the GC as Spitzer mass-segregation instability \citep{spitzer1969} would set in. BHs will strongly interact in this contracting BHS and escape the cluster leaving behind at the most 1 or 2 BHs within a few hundred Myr. However, more recent theoretical and numerical studies
\citep{mackey2008,morscher2013,bh2013a,bh2013b,sh2013,hg2014,ziosi14,morscher2015,wang2016,peuten2016,webb2018,baner2018} have shown that GCs, depending on their initial conditions could sustain a significant population of BHs for several Gyr. These results show that the BHS that forms due to segregating BHs during the early evolution of the GC is not entirely decoupled from the rest of the GC and the evolution of this BHS is governed by the energy demands of its host GC \citep{bh2013a,bh2013b}. If the two-body relaxation time for the GC is sufficiently long, a substantial BHS can survive up to a Hubble time or longer \citep{morscher2015,wang2016,arcasedda2016,rodriguez16,aag}. 

During the last ten years, there have been numerous observational studies that have identified BH candidates in Galactic and
extragalactic GCs \citep{mac2007,barnard1,barnard2,strader2012,roberts2012,chomiuk2013,mj2015,minn2015,bahramian2017}. Kinematic observations of extragalactic GCs have also indicated presence of significant unseen mass \citep{taylor2015}. Most recently, \citet{giesers2018} identified a stellar mass BH
candidate in a detached binary system with a main sequence star in the Galactic GC NGC 3201 using MUSE. 
There have also been recent works that use numerical simulations of GC models harbouring a large population of BHs in order to find signatures for the presence of BHS in GCs \citep{arcasedda2016,weatherford2017,aag}.

\defcitealias{aag}{AAG 2018}

\citet[][hereafter AAG 2018]{aag} proposed a novel way to define the BHS size as the radius within which half-mass is in BHs and the other  half is in stars. Following this definition, \citetalias{aag} made use of more than 150 simulated GC models to find correlations between observational and structural
properties of a GC at 12 Gyr with the properties of the BHS that it
contains at its center. Using those correlations and the observed
properties of Galactic GCs taken from the \citet[][updated 2010]
{Harris1996} catalogue, we identify 29 GCs that are likely
to be harbouring a BHS at their center. We also identify GCs that are
likely to have a significant number of their BHs in binary systems. 
In Section \ref{mocca-models}, we briefly describe the \textsc{mocca} code for star cluster simulations and the GC models that were used in this study. In Section \ref{sec2:-cor}, we describe the correlations that were used to estimate BHS properties from available observational data (all the relations and the fitting parameters are provided in Table 
\ref {tab:app-table} in Appendix \ref{app}). The correlations are also applied to results from \textit{N}-body simulations to see if we can recover BH numbers in those models (Section \ref{test-dragon}). We also give a description for the criteria that was used to shortlist the 29 candidate Galactic GCs with a BHS and discuss the limitations of our work in Sections \ref{criterion} and \ref{limits}. In Section \ref{result},
we provide the observed properties of the 29 Galactic GCs that we have shortlisted in Table \ref{table-obs}. The main result of the paper is shown in Table \ref{table-bhs} where we provide the estimated density, size, mass and average mass of the BHS these 29 Galactic GCs may contain. The total number of BHs and the potential number of BHs in binary systems within these GCs is also estimated. In Section \ref{last-section}, we discuss the results and provide the main conclusions.

\section{Method}\label{sec2:method}

\subsection{MOCCA \& GC Models}\label{mocca-models}

In order to identify which Galactic GCs 
could contain a substantial number of BHs, we made use of a large suite of GC models that were
simulated with the \textsc{mocca} code \citep[see][and reference
therein for details]{hypki2013,giersz2013} for evolving star clusters 
as part of the \textsc{mocca-survey} Database I project
\citep{askar2017}. \textsc{mocca} simulates the dynamical evolution of
a GC based on H{\'e}non's Monte-Carlo algorithm \citep{henon1971} and
improvements to this method \citep{stodolkiewicz1986,giersz1998}.
\textsc{mocca} uses the \textsc{fewbody} code \citep{fewbody} for
computing the outcome of strong interactions, and for binary/stellar
evolution it employs the \textsc{SSE}/\textsc{BSE} codes
\citep{sse,bse}. For the Galactic potential, \textsc{mocca} uses a
simple point mass approximation with the Galaxy mass equal to the mass
enclosed within the Galactocentric distance of the GC model.

The \textsc{mocca-survey} Database I is a collection of about 2000 GC
models that were simulated with the \textsc{mocca} code. These GCs span
a wide array of initial conditions that cover different initial
masses, metallicities, primordial binary fraction, concentration,
half-mass radii, tidal radii and prescriptions for BH natal kicks
\citep[for details see][]{askar2017}. The results presented in this
paper come from 163 GC models that had a population of 15 to
800 BHs at 12 Gyr. These models emerged from all initial
metallicities that were sampled and initial half mass radii for most of these models were 2.4 pc and 4.8 pc.  86 per cent of these models had initial number of stars larger than about $7 \times 10^{5}$ . The details of the simulation models that host a large number of BHs are provided in \citetalias{aag}. For all these GC models, BH natal kicks were modified according to the mass fallback prescription by \citep{bkb2002}. With these natal kicks, BH retention fraction\footnote{This is calculated by taking the number of BHs that are retained in the GC (from evolution of massive stars) divided by the total number of massive stars in the initial model that should evolve into BHs.} within the first 20 -- 30 Myr of evolution ranged from 15 per cent to 55 per cent depending on the initial cluster mass, concentration and tidal radius.       

\subsection{Estimating BHS Properties for Galactic GCs} \label{sec2:-cor}

\citetalias{aag} carried out a detailed investigation of GC models
from the \textsc{mocca-survey} Database I that retained a sizeable
population of BHs at 12 Gyr. Typically, such GC models are
characterized by low central surface brightness (CSB), large half-light radii and present-day half-mass relaxation times larger than a Gyr.
\citetalias{aag} defined the size of the BHS ($R_\bhs$) as the
radius within which 50 per cent of the total cumulative mass is in BHs while the remaining mass is in other stars. Investigating the
properties of the GCs at 12 Gyr, \citetalias{aag} found a tight
correlation between the average GC surface brightness inside the projected
half-light radius ($\rm L_{\rm GC}/ r_{hl}^{2}$ \footnote{$\rm L_{\rm GC}
r_{hl}^{-2}$ is defined as the GC luminosity divided by the square of
the half-light radius ($\rm L_{\odot} \rm pc^{-2}$).}) and the mass density of
the BHS ($\rho^{}_{\bhs}$\footnote{$\rho^{}_{\bhs}$ is defined as the
total mass of BHs inside the BHS divided by the cube of the radius
($R_\bhs$) that defines the size of the BHS ($\rm M_{\odot} \rm
pc^{-3}$).}). Additional correlations were found by \citetalias{aag}
between the BHS density and its radius and mass. Using those correlations, it is possible to estimate the properties of the BHS for an observed GC using its luminosity and half-light radius.

\begin{figure}
	\includegraphics[width=\columnwidth]{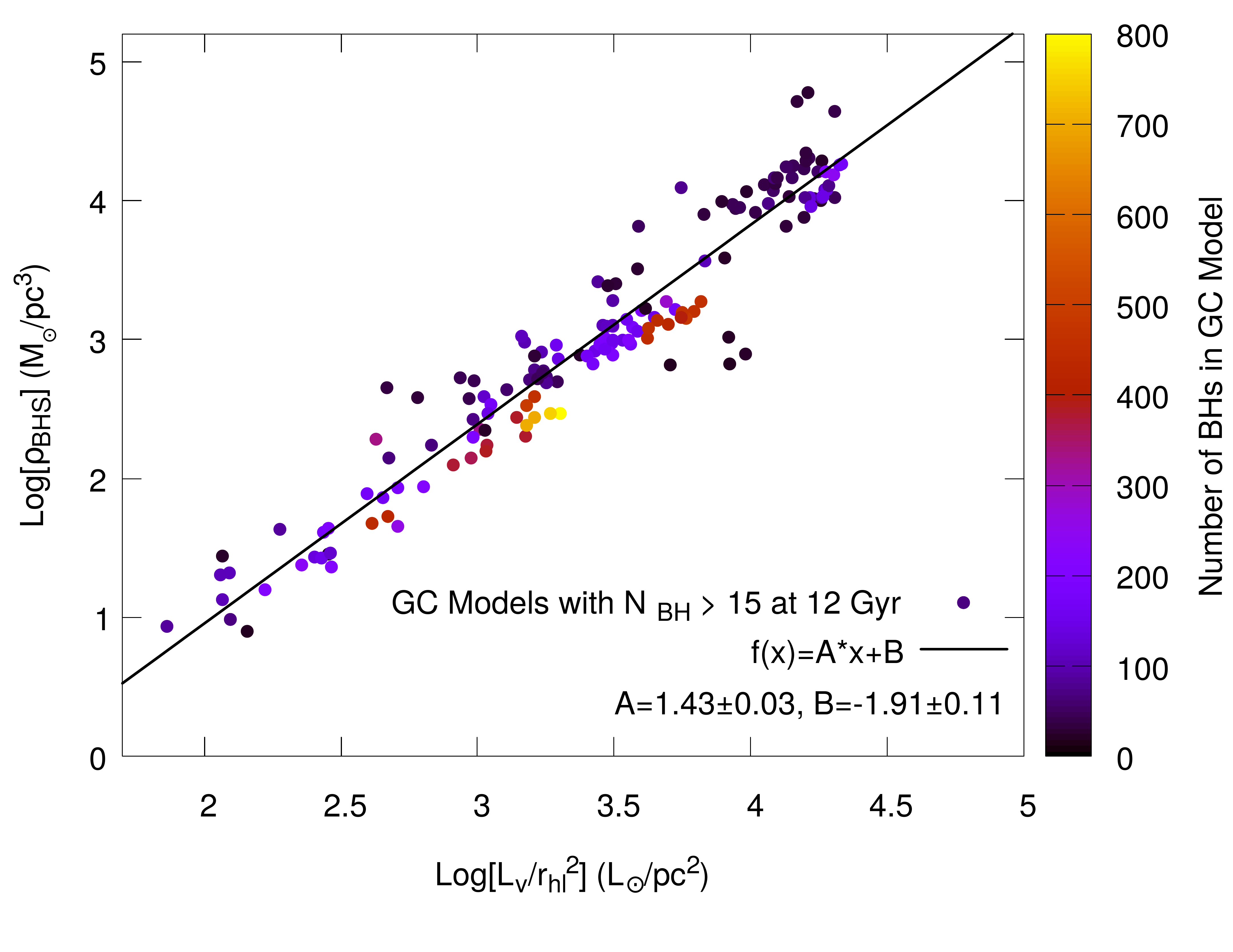}
    \caption{
BHS density as a function of the GC V-band luminosity inside the square of the half-light radius ($L_{\rm V}/r_{hl}^{2}$) at 12 Gyr. The black line shows the fitted linear in log-log correlation. The colours of the points indicate the total number of BHs in the GC model.}
\label{correlation}  
\end{figure}

We apply the correlations found by \citetalias{aag} to the available
observed data for Galactic GCs taken from \citet[][updated 2010
]{Harris1996} which provides half-light radius and absolute V-band
magnitude ($M_{\rm V})$ for 140 Galactic GCs. $M_{\rm V}$ can be
converted to the V-band luminosity of the cluster using the distance
of the GC from the Sun. In Equation 15 of \citetalias{aag}, the
linear in log-log correlation between $L_{\rm GC}/ r_{hl}^{2}$ and
$\rho^{}_{\bhs}$ is provided along with fitted values of the
constants. In that equation, $L_{\rm GC}$ was the total bolometric
luminosity of the GC model at 12 Gyr. In order to apply the correlation
to the available observational data in the Harris catalogue, we
refitted the parameters for the correlation by using the total V-band luminosity ($L_{\rm V}$) of the \textsc{mocca} 163 GC models. Using the absolute V-band magnitude of each
object in the simulation snapshot at 12 Gyr, we converted this
magnitude to luminosity and obtained the total cluster luminosity in
the V-band at 12 Gyr. In Figure \ref{correlation}, we show how $L_{\rm
V}/ r_{hl}^{2}$ correlates with the $\rho^{}_{\bhs}$ at 12 Gyr for
MOCCA models that contained more than 15 BHs at 12 Gyr. 
This correlation is well described by the following relation

\begin{equation}
\Log \left(\frac{\rho^{}_\bhs}{\Ms {\rm pc}^{-3}}\right) = A\left[\Log \left(\frac{L_{\rm V}}{{\rm L}_\odot}\right)-2\Log \left(\frac{r_{\rm hl}}{{\rm pc}}\right)\right]+B,
\label{fun}
\end{equation}
where $A = 1.43\pm 0.03$ and $B = -1.91\pm0.11$
 
Knowing the observed  V-band luminosity and the half-light radius of a Galactic GC from the Harris catalogue, we can estimate the 
density of its BHS ($\rho^{}_\bhs$). \citetalias{aag} found that there is an anti-correlation between the $\rho^{}_\bhs$ and the size of the BHS ($R_\bhs$) \citepalias[][ Equation 6]{aag}.
$R_\bhs$ correlates with $M_\bhs$ (which is the mass of BHs enclosed within $R_\bhs$) and the average mass of BHs within $R_\bhs$
\citepalias[][ Equations 5 and 8]{aag}. $M_\bhs$ correlates with the number of BHs in the subsystem \citepalias[][ Equation 7]{aag}. Moreover, \citetalias{aag} found that $M_\bhs$ and the number of BHs in the BHS correlate with the total mass of all BHs and total number
of all BHs in the GC model. Furthermore, the fraction of BHs in binary systems anti-correlates with the $M_\bhs$ and the average mass of other stars in the BHS also anti-correlates with the average mass of
BHs in the BHS (see Table \ref{tab:app-table} in Appendix \ref{app} for a full list of all the correlated parameters, the correlations and the values of the fitted parameters). All these correlations were exploited to get properties of BHS and over all BH populations for Galactic GCs for 
which we had observed values of V-band luminosity and half-light 
radius.

\subsection{Testing Correlations on Results from \textit{N}-Body Simulations}\label{test-dragon}

\citet{wang2016} carried out direct \textit{N}-body simulations of 
million body GCs using the \textsc{nbody6++gpu} code 
\citep{wang2015} as part of the \textsc{dragon} simulation project. 
The two models that were simulated up to 12 Gyr, D1-R7-IMF93 and D2-R7-IMF01 retained 245 and 1037 BHs at 12 Gyr respectively. We tested 
our correlations to see if we could correctly recover the total number 
of BHs at 12 Gyr in the two \textsc{dragon} clusters by using the 
total V-band luminosity and half-light radius for D1-R7-IMF93 and D2-
R7-IMF01 in \citet{wang2016} .

The half-light radius of model D1-R7-IMF93 at 12 Gyr was $\sim 8.7$ 
pc and its total V-band luminosity was $1.86 \times 10^{5}$ $\rm L_{\odot}$ 
\footnote{Value for V-band luminosity for \textsc{dragon} simulation 
models D1-R7-IMF93 and D2-R7-IMF01 were obtained from the 12 Gyr 
simulation snapshots.}. Using the correlation described in Equation 
\ref{fun}, 
we find Log[$\rho^{}_\bhs$]$= 2.94$. Connecting $\rho_\bhs$ to the BHS size and total mass, we estimate the total number of retained BHs at 12 Gyr in D1-R7-IMF93 to be $N_{\rm BH} = 140^{+80}_{-45}$. Our correlation predicts about 100 
fewer BHs in D1-R7-IMF93. However, the upper limit error limit is off 
by about 25 BHs. The reason for this discrepancy can be attributed to 
the fact that the majority (80 per cent) of the GCs that we considered when 
deriving the correlation in Equation \ref{fun} had between 15 to 200 
BHs at 12 Gyr (see colour bar in Figure \ref{correlation}). The half-light radii at 12 Gyr for more than 60 per cent of 
these GCs was between 2.0 and 6.5 pc. For GCs with a large number of BHs, large half-light radius and a large total luminosity value, the correlation 
in Equation \ref{fun} gives an overestimated value of the density of the BHS which, in turn, reduces the estimated BHS size, mass and estimated total number of retained BHs. GC models retaining several hundreds of BHs 
have large initial half-mass radii and relaxation times thus they are dynamically younger compared to models that retain about several tens to few hundreds of BHs. 

If the fitted parameters in the correlation given in Equation \ref{fun} are 
calculated for 12 GC models in which there are more than 250 BHs, V-band 
luminosity is larger than $1.0 \times 10^{5} \rm L_{\odot}$ and half-light radius is larger than 5 pc, we find A and B to be $1.41 \pm 0.11$ and 
$-2.05 \pm 0.04$ respectively. These values yield Log[$\rho^{}_\bhs]=2.69$, and total number of BHs at 12 Gyr to be 
$N_{\rm BH}$ = $161^{+126}_{-61}$ in D1-R7-IMF93 . 
 The use of 
these fitted parameters can possibly provide an estimate for BHS properties in luminous GCs with large half-light radii.  
For model D2-R7-IMF01, the total V-band luminosity at 12 Gyr was $1.11 \times 10^{5}$ $\rm L_{\rm \odot}$ and the half-light radius was $\sim 14.4$ pc. 

Using Equation \ref{fun} with A = $1.41 \pm 0.11$ and B = $-2.05 \pm 0.04$, 
we find that the Log[$\rho^{}_\bhs$]=1.98. This results in a total number of BHs to be $N_\bh=271^{+251}_{-116}$. Even taking into 
account maximum errors in the correlation, the BHs in D2-R7-IMF01 is 
underestimated by half. 

Fitting values of A and B for 4 MOCCA models which had more than 500 BHs at 12 Gyr gives a correlation where $A = 0.6 \pm 0.2$  and $B = 0.4 \pm 0.7$. Using these values, the maximum number of BHs estimated in D2-R7-IMF01 is around 700. Therefore, it seems that the application of the correlations discussed in this paper to GCs with half-light 
radii larger than 8 pc and above may lead to an underestimation of the number of BHs in those GCs. However, GCs with such large half-light radius values are 
rare in the Galaxy, as there are only 8 Galactic GCs within 50 kpc of the Galactic center which have such a high half-light radii. The 
majority of
Galactic GCs have present-day half-light radii between 2 to 5 pc.

\subsection{GC Identification Criteria}\label{criterion}

By using Equation \ref{fun} and the correlations described in Section 
\ref{sec2:-cor} and \citetalias{aag}, we estimated the BHS properties
and the potential number of BHs in a GC from the observed half-light radius and total V-band luminosity of the GC. We have shortlisted 29 
Galactic GCs for which central surface brightness, absolute magnitude 
and average surface luminosity values agree with what we find for simulated GCs that 
harbour a high number of BHs at 12 Gyr. We have also selected only those 
GCs for which the observed present-day half-mass relaxation times are 
larger than about 0.9 Gyr. This choice is motivated by the properties of our simulations, as all the GC models 
with a high number of BHs had half-mass relaxation times larger than this limiting value at 12 Gyr. 

\begin{figure}
	\includegraphics[width=\columnwidth]{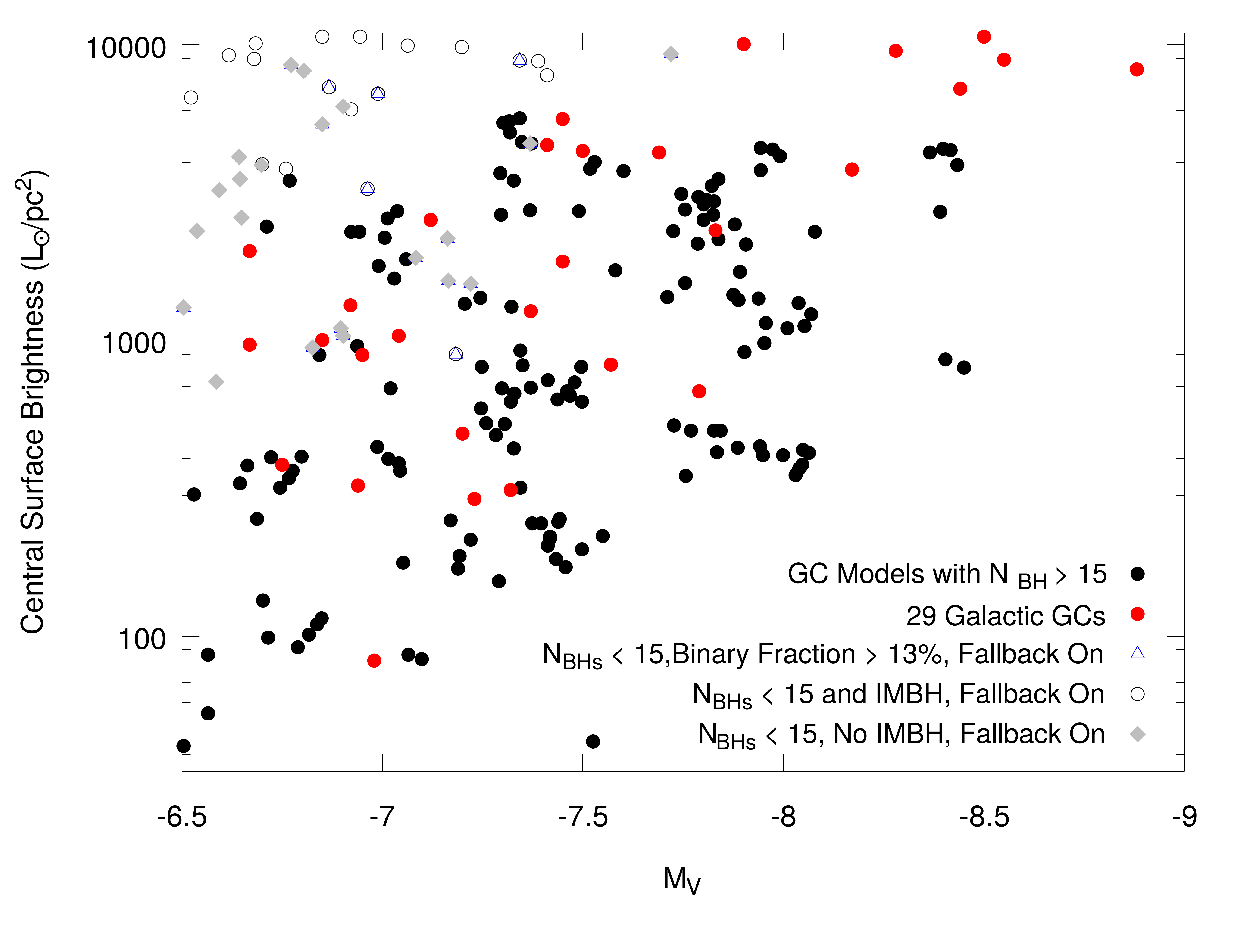}
	\includegraphics[width=\columnwidth]{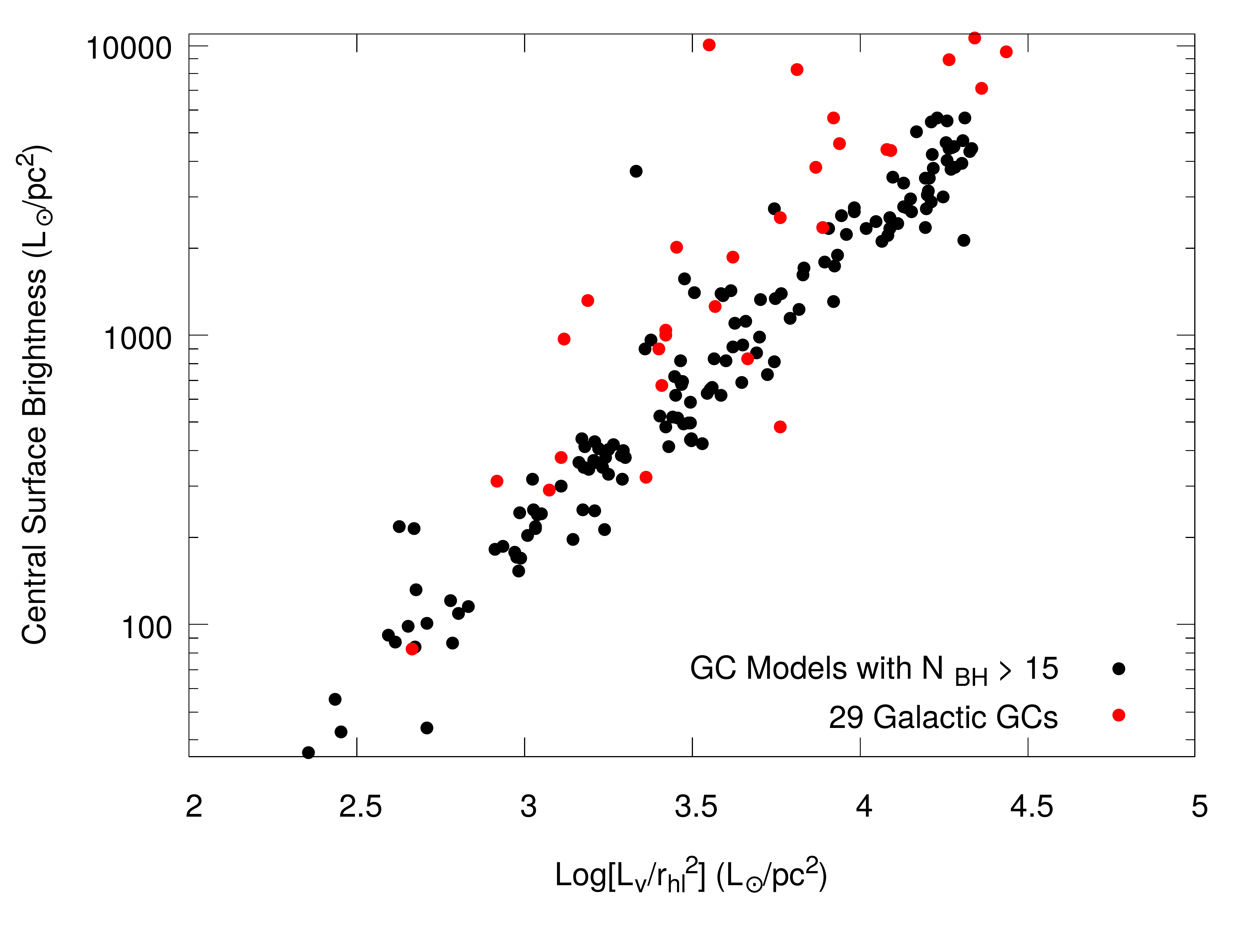}
    \caption{\textbf{Top Panel}: On the x-axis is the absolute V-band magnitude and on the y-axis is the central V-band surface brightness (CSB) in units of $\rm L_{\odot}$ $\rm pc{^-2}$. 12 Gyr properties for simulated GC models with more than 15 BHs are shown with the black points and the observed properties of 29 Galactic GCs that could potentially be containing BHs are shown with red points. Other simulation models which occupy the range of magnitudes and CSB at 12 Gyr are also shown. These are only the models in which mass fallback was enabled but number of BHs at 12 Gyr was less than 15. \textbf{Bottom Panel:} On the x-axis is the GC V-band luminosity divided by the square of the the half-light radius and on the y-axis is the CSB. The colour coding is similar to the top panel.}
\label{fig2}  
\end{figure}

While shortlisting these 29 Galactic GCs, we restricted ourselves to 
GCs for which the Galactocentric radius values were smaller than 17 
kpc. This was done because except for two, all the 163 simulated GC 
models that retained more than 15 BHs at 12 Gyr had Galactocentric radii smaller than 17 kpc. We cannot compare these simulated 
models with Galactic GCs that are very far from the Galactic center 
due to the modeling of the tidal field. Nevertheless, in Section 
\ref{last-section}, we mention distant Galactic GCs that do have 
observational properties suggesting the presence of a large number of 
BHs. Moreover, we shortlisted Galactic GCs which were bright ($M_{\rm V} 
<-6.5$) but had low central surface brightness (CSB $ \lesssim 1 
\times 10^{4}$ $\rm L_{\odot} \rm pc^{-2}$) values similar to the observed properties of simulated GCs at 12 Gyr (see top panel in Figure \ref{fig2}). There were only 3 GC models with BHS that had CSB values larger than $1 \times 10^{4}$ $\rm L_{\odot} \rm pc^{-2}$. In order to calculate the CSB from the simulated models, we use the infinite projection method described in Appendix B of \citet{mashchenko2005} to generate a surface brightness profile for the GC at 12 Gyr, which we use to find the central value
in units of V-band luminosity per square pc. In order to compare this value with the observations, we convert the apparent V magnitudes per square arcsecond CSB value provided in the \citet[][updated 2010]{Harris1996} catalogue to units of V-band luminosity per square pc. In doing this, we took 
into account the distance to the cluster from the Sun and the foreground reddening
are also provided in the catalogue.

We also compared the V-band luminosity inside the half-light 
radius and the central surface brightness of the observed GCs with the 
simulated models to further constrain our list of Galactic GCs with a BHS (see lower panel in Figure \ref{fig2}). We excluded two very bright and massive GCs, these were NGC 5139 ($\rm \omega$ Cen) and NGC 6402 (M14). 
While both these GCs are bright with relatively low central surface 
brightness values, none of the simulated models at 12 Gyr are as bright as these GCs. To reproduce their present-day brightness, we would need initial masses larger than the most massive GC models simulated in the \textsc{mocca-survey} Database I.

The Galactic GC selection criteria discussed above are summarized in the list below:
\begin{itemize}
\item We considered only Galactic GCs with Galactocentric radius values smaller than 17 kpc as we cannot compare our simulated models with very distant GCs.
\item We only 
consider Galactic GCs which were brighter than $M_{\rm V} <-6.5$ with the exception of NGC 5139 ($\rm \omega$ Cen) and NGC 6402 (M14).
\item  We 
restrict ourselves to Galactic GCs having
CSB values $ \lesssim 1 \times 10^{4}$ $\rm L_{\odot} \rm pc^{-2}$.
\item Observed present-day half-mass relaxation time of shortlisted GCs is 
$\gtrsim 0.9$ Gyr.
\end{itemize}

\subsection{Comparison with Central Kinematic Properties}\label{kinematic}
\begin{figure}
	\includegraphics[width=\columnwidth]{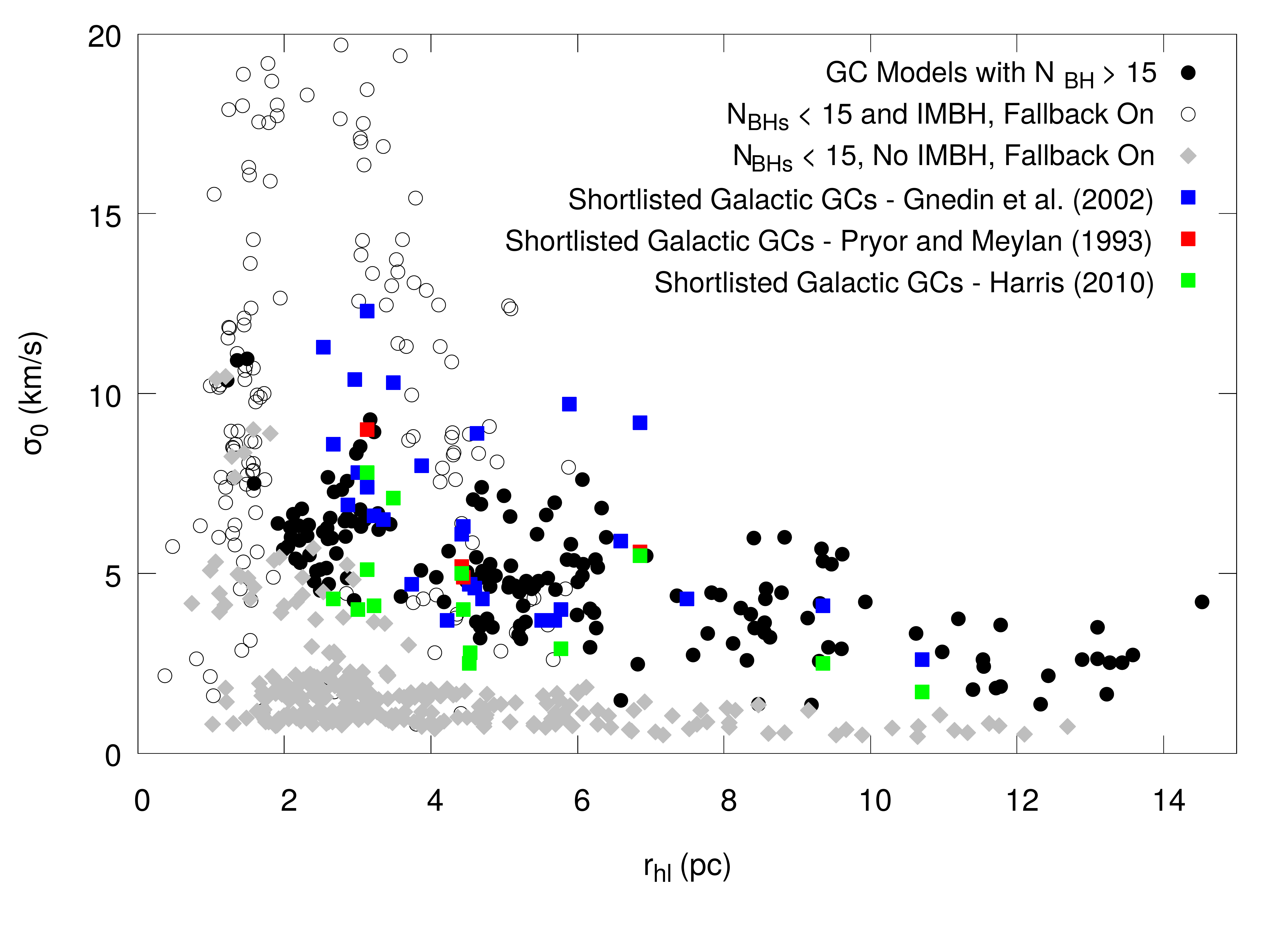}
	\includegraphics[width=\columnwidth]{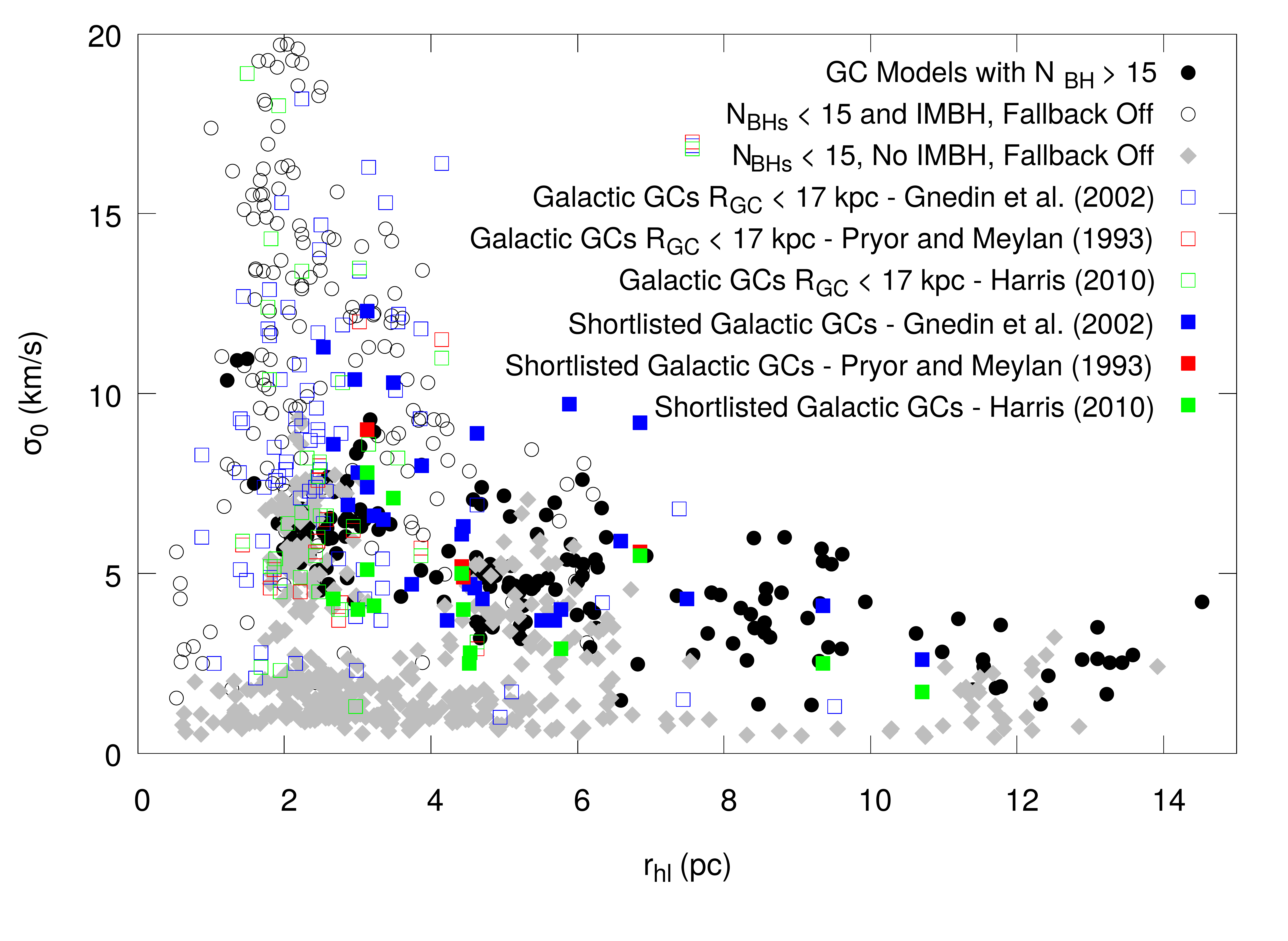}
    \caption{\textbf{Top Panel}: On the x-axis is the half-light radius ($\rm r_{hl}$ in pc and on the y-axis is the central line-of-sight velocity dispersion ($\sigma_{0}$) in units of $\rm km\rm s^{-1}$ at 12 Gyr. Simulated GC models with more than 15 BHs are shown with black points and the available observed values for the shortlisted Galactic GCs from three different sources \citep[][updated 2010]{pryor93,gnedin2002,Harris1996} are shown with squares. Other simulation models in which mass fallback was enabled but the GCs have less than 15 BHs at 12 Gyr are also shown with empty circles (models with an IMBH) and grey diamond points (no IMBH or BHS). Maximum central line-of-sight velocity dispersion values have been limited to 20 $\rm km\rm s^{-1}$ in the figure. Few models with IMBH can have higher central velocity dispersions. \textbf{Bottom Panel:} The same figure as the top panel. However here, the models in which fallback was not enabled are shown and GCs for which velocity dispersion data was available but they were not shortlisted according to our selection criteria are also shown with empty squares.}
\label{fig-kinematic}  
\end{figure}

We also compared the observed central velocity dispersion ($\sigma_{0}$) value for Galactic GCs with the simulated GC models. To obtain the central velocity dispersion from our simulation models, we projected the 12 Gyr simulated GC models and created
a line-of-sight (LOS) velocity dispersion profile (computed in 50 radial logarithmic bins) for stars brighter than $\rm M_{V}=6$ and took the 
velocity dispersion value in the innermost bin. For GC models that contain more than 15 BHs at 12 Gyr, the median central LOS velocity dispersion value is about 4.9 $\rm km\rm s^{-1}$. For models with fallback enabled and no IMBH or significant number of BHs, the median central LOS velocity dispersion is about 2 $\rm km\rm s^{-1}$. In the top panel of Figure \ref{fig-kinematic}, we show the 12 Gyr central LOS velocity dispersion and half-light radii of simulated models with fallback enabled, models with more than 15 BHs are indicated with  
black points, empty circles identify models containing 
an IMBH and grey points are models with neither an IMBH or a significant number of BHs. 

Models with more than 15 BHs have systematically larger central LOS velocity dispersions compared to models with less than 15 BHs and no IMBH. Observational and estimated central LOS velocity dispersions for the shortlisted Galactic GCs from three different references \citep[][updated 2010]{pryor93,gnedin2002,Harris1996} \footnote{Data for velocity dispersions from \citet{pryor93,gnedin2002} can be found online at the following links:\\\url{http://www-personal.umich.edu/~ognedin/gc/pm93_table2.dat} and\\\url{http://www-personal.umich.edu/~ognedin/gc/vesc.dat}} are shown with filled squares. Two of the shortlisted GCs (NGC 288 and M22) also had LOS velocity dispersion profiles provided in \citet{watkins2015} and central values were similar to those
available in \citet[][updated 2010]{Harris1996}. Observational estimates for central LOS velocity dispersions are not available for all Galactic GCs and values can vary between different studies. However, the shortlisted GCs have central LOS velocity dispersions that are in better agreement with models having a higher number of stellar mass BHs compared to other models in which fallback was enabled but there were no significant number of BHs at 12 Gyr. 

\subsection{Limitations and Cautions}\label{limits}

\begin{figure}
	\includegraphics[width=\columnwidth]{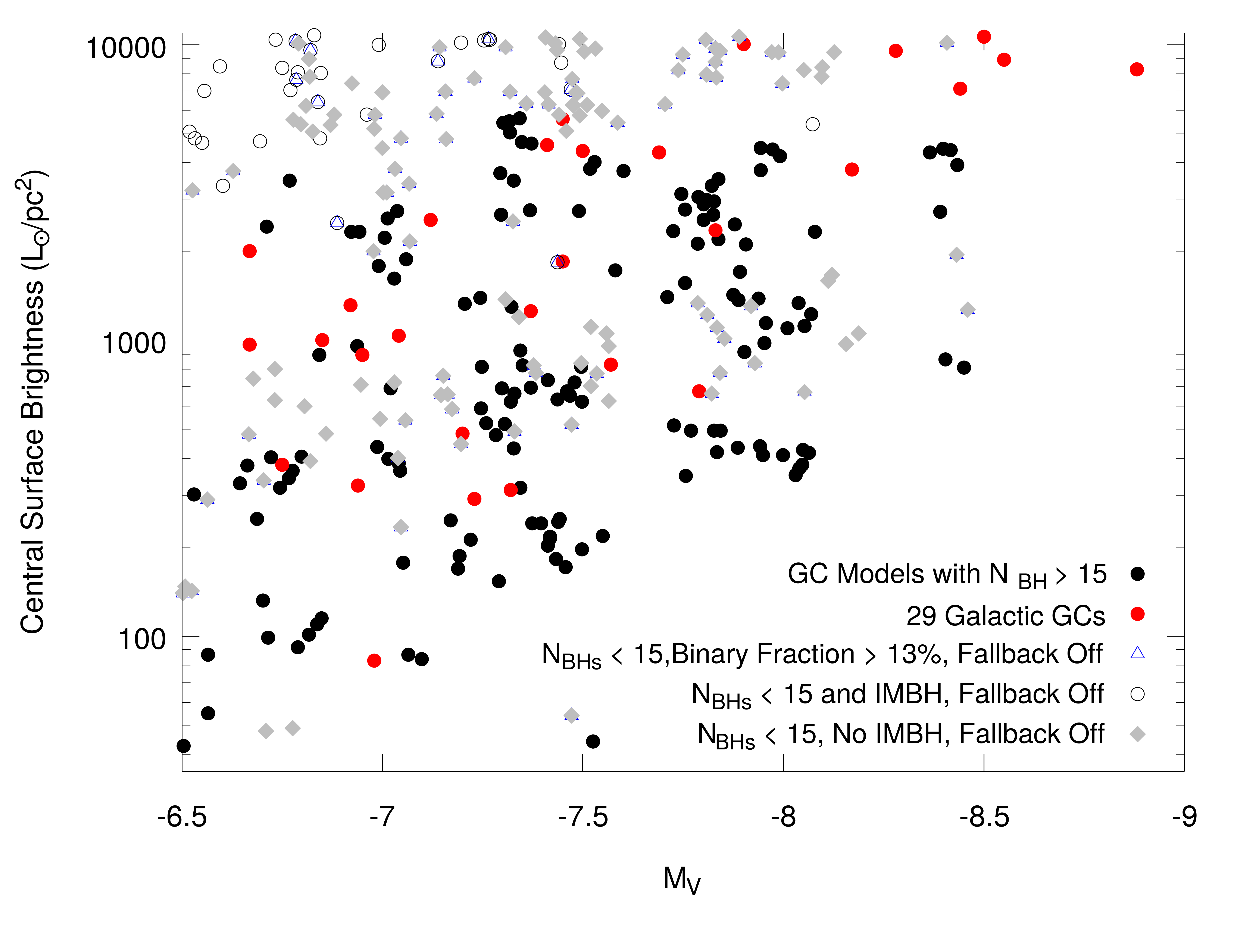}
    \caption{On the x-axis is the absolute V-band magnitude and on the y-axis is the central V-band surface brightness. The figure is the same plot as in the top panel of Figure \ref{fig2}). However, here models with fewer than 15 BHs that emerged from initial conditions in which fallback was turned off and BHs had high natal kicks are also shown.}
\label{fig3}  
\end{figure}

There are about two hundred GC models in \textsc{mocca-survey} Database I with a few BHs ($\lesssim 15$) that at 12 Gyr have total V-band luminosity, central surface brightness and luminosity inside the square of the half-light radius values comparable to models that have
a large number of BHs . More than 80 per cent of these models emerge out of simulations in which \citet{bkb2002} mass fallback prescription was not used (see Figure \ref{fig3}) and BHs had high natal kicks\footnote{\label{note1}Fallback prescription is 
turned off in these GC simulations and BHs get same natal kicks as neutron stars. Distribution of kick velocities is given by a Maxwellian distribution with $\sigma=265$ $\rm km$  $\rm s^{-1}$ \citep{hobbs2005}.}.
Based on these findings, it is important to stress that the results obtained in this paper depend significantly on the prescriptions for BH natal kicks. If BH natal kicks are as high as the natal kicks for neutron stars, inferred from proper motions of pulsars \citep{hobbs2005}, then the observational properties of the 29 shortlisted GCs could still be explained without invoking the need for a BHS. However, if BH natal kicks are lower in low metallicity environments \citep{bkb2002, kb2010, fryer2012,spera2017} like GCs and retention fractions are as high as 50 per cent, then it would be difficult to explain the present-day observational properties of these GCs without the presence of a sizeable number of BHs at 12 Gyr.

For the nearly two hundred GC models for which we assumed high natal kicks for BHs, about half of these GCs are characterized by a high overall binary fraction (between 11 per cent and 37 per cent) at 12 Gyr with a mean value of 20 per cent.
Some of these models even have central velocity dispersion values that are in agreement with observed values of central velocity dispersion for few of our shortlisted Galactic GCs (as shown by the grey points in the lower panel of Figure \ref{fig-kinematic}).
It may be possible that among the 29 short-listed GCs, a high fraction of binaries (about 20 per cent) could explain the observed low central surface brightness and large half-light radii instead of the presence of a BHS. The remaining half of low BH GC models can be split into two types. About 50 of those GCs contain an intermediate mass BH (IMBH). The IMBH in these GCs has typically a low mass, ranging from about a few hundred to a few thousand solar masses. Such IMBHs mostly form via the slow scenario described in \citet{giersz2015} and gradually build up mass. In a few GC models, an IMBH may form quickly due to high central concentrations from runaway collisions in the very early evolution of the clusters, but its subsequent growth slows down due to the expansion of the GC driven by mass loss due to stellar evolution. For another 50 GCs, there are no significant number of BHs, IMBH or high binary fractions. Most of these GCs have large tidal radii and mean initial half-mass radii of about 4 pc. Initial half-mass relaxation times are of the order of several hundred Myr to Gyr. As BH natal kicks are large, such models lose a lot of mass in their early evolution and initially expand but start to gradually collapse after a few or several Gyr of evolution (depending on initial half-mass relaxation times). While half-mass radius keeps expanding, such GCs are typically characterized by low observed core radius values at 12 Gyr of about less than a 1 pc. If BHs 
get high natal kicks at birth then some of the GCs shortlisted as 
harbouring a BHS in this paper may in fact, be GCs with large 
tidal radii that are evolving towards core collapse at the present time.

It is important to point out that the correlations obtained by 
\citetalias{aag} used 12 Gyr snapshots of simulated cluster models 
with different fixed initial metallicities whereas Galactic GCs show an age spread (with typical ages being about 10-13 Gyr) and 
metallicity spread. While this can slightly change the estimated 
values for the properties of the BHS and the numbers presented in this 
paper, we do not expect large differences as GCs that keep a large 
number of BHs up to larger times are dynamically younger with large 
half-mass relaxation times. If the Galactic GCs that we have 
identified are younger than 12 Gyr then they could be retaining a
slightly larger number of BHs, so the estimated number of BHs for GCs 
that are younger than 12 Gyr can be seen as a lower limit.

The correlations applied in this paper make use of the observed 
magnitude and half-light radius values provided in the \citet[]
[updated 2010]{Harris1996} catalogue. While we have used the errors in 
the correlations to calculate maximum and minimum values for the 
inferred properties of the BHS, the errors for these two observed 
parameters are not taken into account and exact values are taken from the Harris catalogue. The luminosity and the half-light radius of the GC depend on the total magnitude and the estimated distance to the 
GC. If errors are large in these two observed values then this can further change the estimated values we obtained for the
various properties of the BHS and overall BH population of the 
clusters. One approach to minimize observational uncertainties 
would be to take average values for parameters like absolute magnitude and distance to the GC from different observational studies. This was done by \citet{baumgardt2017} when comparing results from \textit{N}-body simulations with Galactic GCs. However, this can be done for only a limited number of Galactic GCs. With better and more constrained observations of these global GC properties, such errors can be taken into account in future studies in order to exhaustively establish a list of GCs that could have a large population of unseen BHs.

\section{Results: Galactic Globular Clusters with BHS}\label{result}

\begin{figure}
	\includegraphics[width=\columnwidth]{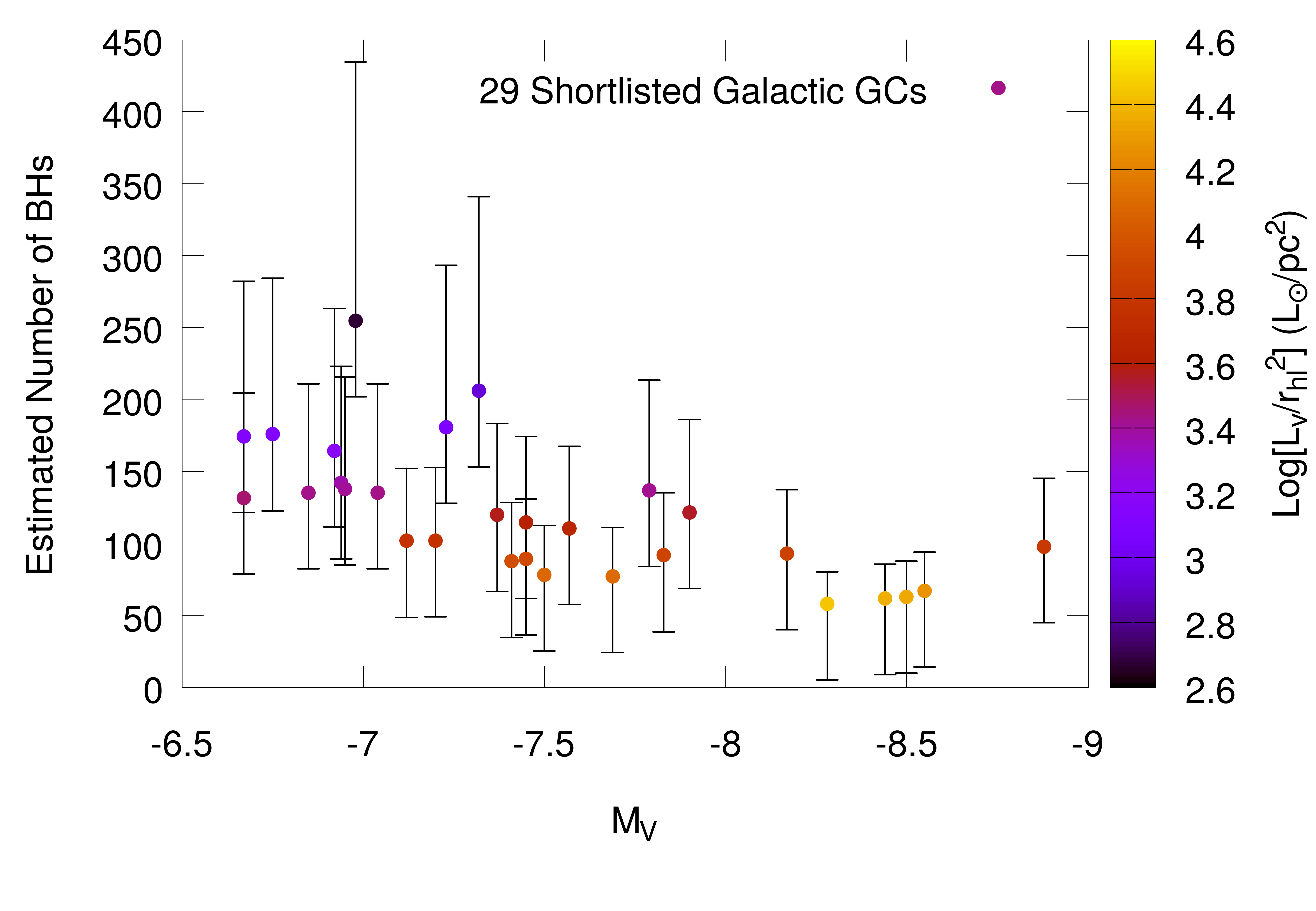}
    \caption{Figure shows the absolute V-band magnitude of the 29 shortlisted Galactic GCs and the total estimated number of BHs they could be harbouring. The colour of the points indicates the average surface brightness (Log$[\rm L_{\rm V}/r_{hl}^{2}]$) obtained from the parameters provided in the \citet[]
[updated 2010]{Harris1996} catalogue.}
\label{fig4}  
\end{figure}

The observed properties of the 29 Galactic GCs that could contain a BHS are provided in Table \ref{table-obs}. These values were 
taken from the \citet[][updated 2010]{Harris1996} catalogue. For 
certain values such as core radius, half-light radius and central 
surface brightness, the units from Harris catalogue were converted to 
more suitable units (see notes of Table \ref{table-obs}) for applying the correlation in Equation \ref{fun} . We used the observed Log$[\rm L_{\rm V}/r_{hl}^{2}]$ value to estimate the
density of the BHS using Equation \ref{fun}. We used subsequent correlations described in Section \ref{sec2:-cor} and Appendix \ref{app} to obtain values for
the properties of the BHS, number of BHs in the GC and potential number of BHs in the binary systems (which includes binary BHs and BHs in binary systems with other stars). All these values are provided in Table \ref{table-bhs}. Figure \ref {fig4} shows the absolute V-band magnitude and the total estimated number of BHs in these 29 Galactic GCs. The colour bar in the figure shows the observed Log$[\rm L_{\rm V}/r_{hl}^{2}]$ for the GCs. 

\begin{table*}
\renewcommand{\arraystretch}{1.1}
\setlength{\tabcolsep}{3.9pt}
\begin{threeparttable}
 \caption{Observational properties \citep[][updated 2010]{Harris1996} of 29 Galactic GCs that are likely to be containing a subsystem of BHs.}
\label{table-obs}
\begin{tabular}{rrrrrrrrcc}
\hline
\multicolumn{1}{c}{\textbf{\begin{tabular}[c]{@{}c@{}}GC\\  Name\end{tabular}}} & \multicolumn{1}{c}{\textbf{$M_{\rm V}$}\tnote{a}} & \multicolumn{1}{c}{\textbf{\begin{tabular}[c]{@{}c@{}}$D_{\rm Sun}$\\ (kpc)\end{tabular}}} & \multicolumn{1}{c}{\textbf{\begin{tabular}[c]{@{}c@{}}$D_{\rm GC}$\\ (kpc)\end{tabular}}} & \multicolumn{1}{c}{\textbf{\begin{tabular}[c]{@{}c@{}}$t_{rh}$\\  (Myr)\end{tabular}}\tnote{b}} & \multicolumn{1}{c}{\textbf{\begin{tabular}[c]{@{}c@{}}$L_{\rm V}$\\ ($\rm L_{\odot}$)\end{tabular}}\tnote{c}} & \multicolumn{1}{c}{\textbf{\begin{tabular}[c]{@{}c@{}}$r_c$\\ (pc)\end{tabular}}} & \multicolumn{1}{c}{\textbf{\begin{tabular}[c]{@{}c@{}}$r_{hl}$\\ (pc)\end{tabular}}} & \multicolumn{1}{c}{\textbf{\begin{tabular}[c]{@{}c@{}}Log{[}CSB{]}\\ $(\rm L_{\odot} \rm pc^{-2})$\end{tabular}}\tnote{d}} & \multicolumn{1}{c}{\textbf{\begin{tabular}[c]{@{}c@{}}Log$[L_{\rm V}/r_{hl}^{2}]$\\ $(\rm L_{\odot} \rm pc^{-2})$\end{tabular}}} \\ \hline
NGC 288 & -6.75 & 8.9 & 12.0 & 2089.3 & $4.29 \times 10^{4}$ & 3.50 & 5.77 & 2.58 & 3.11 \\
NGC 3201 & -7.45 & 4.9 & 8.8 & 1862.1 & $8.17 \times 10^{4}$ & 1.85 & 4.42 & 3.27 & 3.62 \\
NGC 4372 & -7.79 & 5.8 & 7.1 & 3890.5 & $1.12 \times 10^{5}$ & 2.95 & 6.60 & 2.83 & 3.41 \\
\begin{tabular}[c]{@{}r@{}}NGC 4590\\ (M68)\end{tabular} & -7.37 & 10.3 & 10.2 & 1862.1 & $7.59 \times 10^{4}$ & 1.74 & 4.52 & 3.10 & 3.57 \\
NGC 4833 & -8.17 & 6.6 & 7.0 & 2630.3 & $1.58 \times 10^{5}$ & 1.92 & 4.63 & 3.58 & 3.87 \\
\begin{tabular}[c]{@{}r@{}}NGC 5272\\ (M3)\end{tabular} & -8.88 & 10.2 & 12.0 & 6166.0 & $3.05 \times 10^{5}$ & 1.10 & 6.85 & 3.92 & 3.81 \\
NGC 5466 & -6.98 & 16.0 & 16.3 & 5754.4 & $5.30 \times 10^{4}$ & 6.66 & 10.70 & 1.92 & 2.66 \\
IC 4499 & -7.32 & 18.8 & 15.7 & 5370.3 & $7.24 \times 10^{4}$ & 4.59 & 9.35 & 2.50 & 2.92 \\
NGC 5897 & -7.23 & 12.5 & 7.4 & 3715.4 & $6.67 \times 10^{4}$ & 5.09 & 7.49 & 2.46 & 3.07 \\
NGC 5986 & -8.44 & 10.4 & 4.8 & 1513.6 & $2.03 \times 10^{5}$ & 1.42 & 2.96 & 3.85 & 4.36 \\
NGC 6101 & -6.94 & 15.4 & 11.2 & 1659.6 & $5.11 \times 10^{4}$ & 4.35 & 4.70 & 2.51 & 3.36 \\
NGC 6144 & -6.85 & 8.9 & 2.7 & 1380.4 & $4.70 \times 10^{4}$ & 2.43 & 4.22 & 3.00 & 3.42 \\
\begin{tabular}[c]{@{}r@{}}NGC 6171\\ (M107)\end{tabular} & -7.12 & 6.4 & 3.3 & 1000.0 & $6.03 \times 10^{4}$ & 1.04 & 3.22 & 3.40 & 3.76 \\
\begin{tabular}[c]{@{}r@{}}NGC 6205\\ (M13)\end{tabular} & -8.55 & 7.1 & 8.4 & 1995.62 & $2.25 \times 10^{5}$ & 1.28 & 3.49 & 3.95 & 4.26 \\
NGC 6362 & -6.95 & 7.6 & 5.1 & 1584.9 & $5.15 \times 10^{4}$ & 2.50 & 4.53 & 2.95 & 3.40 \\
NGC 6401 & -7.9 & 10.6 & 2.7 & 3388.4 & $1.24 \times 10^{5}$ & 0.77 & 5.89 & 4.00 & 3.55 \\
NGC 6426 & -6.67 & 20.6 & 14.4 & 1905.5 & $3.98 \times 10^{4}$ & 1.56 & 5.51 & 2.99 & 3.12 \\
NGC 6496 & -7.2 & 11.3 & 4.2 & 1096.5 & $6.49 \times 10^{4}$ & 3.12 & 3.35 & 2.68 & 3.76 \\
\begin{tabular}[c]{@{}r@{}}IC 1276\\ (Pal 7)\end{tabular} & -6.67 & 5.4 & 3.7 & 1071.5 & $3.98 \times 10^{4}$ & 1.59 & 3.74 & 3.30 & 3.46 \\
NGC 6569 & -8.28 & 10.9 & 3.1 & 1122.0 & $1.75 \times 10^{5}$ & 1.11 & 2.54 & 3.98 & 4.44 \\
NGC 6584 & -7.69 & 13.5 & 7.0 & 1047.1 & $1.02 \times 10^{5}$ & 1.02 & 2.87 & 3.64 & 4.09 \\
\begin{tabular}[c]{@{}r@{}}NGC 6656\\ (M22)\end{tabular} & -8.5 & 3.2 & 4.9 & 1698.2 & $2.15 \times 10^{5}$ & 1.24 & 3.13 & 4.03 & 4.34 \\
NGC 6712 & -7.5 & 6.9 & 3.5 & 891.3 & $8.55 \times 10^{4}$ & 1.53 & 2.67 & 3.64 & 4.08 \\
NGC 6723 & -7.83 & 8.7 & 2.6 & 1737.8 & $1.16 \times 10^{5}$ & 2.10 & 3.87 & 3.37 & 3.89 \\
\begin{tabular}[c]{@{}r@{}}NGC 6779\\ (M56)\end{tabular} & -7.41 & 9.4 & 9.2 & 1023.3 & $7.87 \times 10^{4}$ & 1.20 & 3.01 & 3.66 & 3.94 \\
\begin{tabular}[c]{@{}r@{}}NGC 6809\\ (M55)\end{tabular} & -7.57 & 5.4 & 3.9 & 1949.8 & $9.12 \times 10^{4}$ & 2.83 & 4.45 & 2.92 & 3.66 \\
Pal 11 & -6.92 & 13.4 & 8.2 & 2187.8 & $5.01 \times 10^{4}$ & 4.64 & 5.69 & 3.12 & 3.19 \\
NGC 6934 & -7.45 & 15.6 & 12.8 & 1096.5 & $8.17 \times 10^{4}$ & 1.00 & 3.13 & 3.75 & 3.92 \\
\begin{tabular}[c]{@{}r@{}}NGC 6981\\ (M71)\end{tabular} & -7.04 & 17.0 & 12.9 & 1698.244 & $5.60 \times 10^{4}$ & 2.27 & 4.60 & 3.02 & 3.42 \\ \hline
\end{tabular}
\begin{tablenotes}
       \item[a] Absolute V band magnitude of the GC as provided by  \citet[][updated 2010]{Harris1996}. $D_{sun}$ and $D_{GC}$ are the distance to the cluster from the Sun and the Galactic center in units of kpc. \item[b] Is the current estimated half-mass relaxation time in Myr provided in the Harris catalogue. We restricted our results to models for which this value was greater than about 1 Gyr. \item[c]  Is the luminosity of the GC in V band calculated using the absolute magnitude and the distance to the cluster from the Sun. $r_{c}$ (core radius) and $r_{hl}$ (half-light radius) provided in the Harris catalogue in arcmins were converted to parsec. \item[d] CSB was calculated by converting the apparent V band magnitude per square arcsecond value provided in the Harris catalogue to luminosity per square parsec taking into the account the E(B-V) reddening correction and the distance to the cluster. The last column provides the common logarithm value of the average surface brightness which is defined as the total luminosity divided by the square of the half-light radius.
       
\end{tablenotes}
\end{threeparttable}
\end{table*}

For the 29 selected Galactic GC models, the mean present-day half-mass 
relaxation time is $2.25$ Gyr and the mean half-light radius is 4.93 
pc. The mass of the BHS from the 29 GCs ranges between $\sim 
300$ to $2500 \rm M_{\odot}$. The error bars, particularly upper limits 
for BHS mass and number of BHs, are large. Overall, the total mass of BHs in 
these GCs could be as high as several thousand solar masses The mean mass density of BHs 
in the BHS is about 776 $\rm M_\odot \rm pc^{-3}$. 

As shown from the correlations discussed in our companion paper \citetalias{aag}, models with relatively sparse extended BHS have higher BHS mass and a larger number of BHs. We have also estimated the number of BHs that could be in binary systems in these clusters. In order to obtain this number, we obtained a linear 
anti-correlation that best fits the binary fraction of BH binaries for models with 
BHS mass in the range of 400 to 3000 $\rm M_{\odot}$. While the errors 
on these values are large, as an upper limit we can estimate that 
these GCs can on an average contain from about 8 to a few dozen of binaries with at least one BH component at 12 Gyr. A fraction of these binaries would be detectable mass transferring systems \citep{kremer2018}, these low numbers can explain the dearth of observed BH binaries in GCs that are X-ray or radio sources. There is an anti-correlation between the number of BHs in binary systems and the mass of the BHS (see Figure 4 in \citetalias{aag} for the exact relation). So while GCs with low mass BHS (dynamically older BHS) are more likely to have a higher fraction of their BH population in binaries, the total number of BHs and average mass of BHs would be low for such GCs which explains why we estimate roughly the same number of BH binaries in 29 GCs for which we have a significant range of BHS masses. More massive and dynamically younger BHS have more BHs but as this BHS is less evolved, fewer fraction of BHs are in binares \citepalias{aag}. Our results show that the detection of even a few detached or mass transferring BH binaries in these 29 shortlisted GCs could be an indication of the presence of a larger population of single BHs, particularly if the estimated size of the BHS is large. 

\begin{table*}
\setlength{\tabcolsep}{0.01pt}
\renewcommand{\arraystretch}{1.3}
\begin{threeparttable}

 \caption{Estimated properties of BHS and BH populations in 29 GCs that were shortlisted according the criterion explained in Section \ref{criterion}.}
\label{table-bhs}
\hskip-1.0cm
\begin{tabular}{rcrrrrcrrc}
\hline
\multicolumn{1}{c}{\textbf{\begin{tabular}[c]{@{}c@{}}GC\\  Name\end{tabular}}} & \multicolumn{1}{c}{\textbf{\begin{tabular}[c]{@{}c@{}}$\sim$ Log$[\rho^{}_{\bhs}]$\\ $(\rm M_{\odot} \rm pc^{-3})$\end{tabular}} \tnote{a}} & \multicolumn{1}{c}{\textbf{\begin{tabular}[c]{@{}c@{}}$\sim R_{\bhs}$\\ (pc)\end{tabular}}\tnote{b}} & \multicolumn{1}{c}{\textbf{\begin{tabular}[c]{@{}c@{}}$\sim M_{\bhs}$\\ $(\rm M_{\odot})$\end{tabular}}\tnote{c}} & \multicolumn{1}{c}{\textbf{\begin{tabular}[c]{@{}c@{}}$\sim N_{\bh} $\\ in BHS\end{tabular}}\tnote{d}} & \multicolumn{1}{c}{\textbf{\begin{tabular}[c]{@{}c@{}}Mean $ \sim M_{\bh}$\\  in BHS\\  $(\rm M_{\odot})$\end{tabular}}\tnote{e}} & \multicolumn{1}{c}{\textbf{\begin{tabular}[c]{@{}c@{}}Mean $\sim M_{\star}$\\  in BHS\\  $\pm 0.06(\rm M_{\odot})$\end{tabular}}\tnote{f}} & \multicolumn{1}{c}{\textbf{\begin{tabular}[c]{@{}c@{}}$\sim M_{\bh}$\\  in GC\\ $(\rm M_{\odot})$\end{tabular}}\tnote{g}} & \multicolumn{1}{c}{\textbf{\begin{tabular}[c]{@{}c@{}}$\sim N_{\bh}$\\in GC\end{tabular}}\tnote{h}} & \multicolumn{1}{c}{\textbf{\begin{tabular}[c]{@{}c@{}}$\sim N_{\bh}$\\  in\\Binaries\end{tabular}}\tnote{i}} \\ \hline
NGC 288 & $2.54 \pm 0.20$ & $1.42^{+0.45}_{-0.32}$ & $1473.0^{+566}_{-354}$ & $118^{+58}_{-35}$ & $12.7^{+0.7}_{-0.4}$ & $0.44$ & $2153.6^{+1206}_{-703}$ & $176^{+108}_{-61}$ & $9^{+20}_{-6}$ \\
NGC 3201 & $3.27 \pm 0.22$ & $0.64^{+0.19}_{-0.14}$ & $795.9^{+237}_{-152}$ & $68^{+27}_{-17}$ & $11.4^{+0.5}_{-0.4}$ & $0.49$ & $1275.7^{+596}_{-365}$ & $114^{+60}_{-35}$ & $7^{+15}_{-5}$ \\
NGC 4372 & $2.97 \pm 0.21$ & $0.89^{+0.28}_{-0.20}$ & $1027.0^{+342}_{-217}$ & $85^{+37}_{-23}$ & $11.9^{+0.6}_{-0.5}$ & $0.47$ & $1584.3^{+800}_{-480}$ & $137^{+77}_{-45}$ & $8^{+17}_{-5}$ \\
\begin{tabular}[c]{@{}r@{}}NGC 4590 \\ (M68)\end{tabular} & $3.19 \pm 0.22$ & $0.69^{+0.21}_{-0.15}$ & $847.8^{+260}_{-166}$ & $71^{+29}_{-18}$ & $11.5^{+0.5}_{-0.4}$ & $0.49$ & $1346.0^{+641}_{-391}$ & $120^{+64}_{-37}$ & $8^{+17}_{-5}$ \\
NGC 4833 & $3.62 \pm 0.23$ & $0.43^{+0.13}_{-0.09}$ & $590.8^{+152}_{-98}$ & $52^{+18}_{-12}$ & $10.9^{+0.4}_{-0.3}$ & $0.52$ & $990.2^{+420}_{-263}$ & $93^{+44}_{-27}$ & $8^{+16}_{-5}$ \\
\begin{tabular}[c]{@{}r@{}}NGC 5272\\ (M3)\end{tabular} & $3.54 \pm 0.22$ & $0.47^{+0.14}_{-0.10}$ & $632.9^{+169}_{-109}$ & $55^{+20}_{-13}$ & $11.0^{+0.5}_{-0.4}$ & $0.51$ & $1049.9^{+455}_{-284}$ & $98^{+48}_{-29}$ & $7^{+14}_{-5}$ \\
NGC 5466 & $1.90 \pm 0.19$ & $2.85^{+0.94}_{-0.67}$ & $2512.2^{+1165}_{-703}$ & $191^{+110}_{-63}$ & $13.9^{+0.9}_{-0.7}$ & $0.42$ & $3388.9^{+2188}_{-1220}$ & $255^{+180}_{-96}$ & $10^{+26}_{-7}$ \\
IC 4499 & $2.26 \pm 0.20$ & $1.92^{+0.62}_{-0.44}$ & $1852.8^{+774}_{-477}$ & $145^{+77}_{-45}$ & $13.2^{+0.8}_{-0.6}$ & $0.43$ & $2616.1^{+1560}_{-893}$ & $206^{+135}_{-74}$ & $9^{+22}_{-6}$ \\
NGC 5897 & $2.49 \pm 0.14$ & $1.50^{+0.48}_{-0.34}$ & $1534.7^{+599}_{-373}$ & $122^{+61}_{-36}$ & $12.8^{+0.7}_{-0.6}$ & $0.44$ & $2229.1^{+1263}_{-734}$ & $181^{+113}_{-63}$ & $9^{+21}_{-6}$ \\
NGC 5986 & $4.33 \pm 0.24$ & $0.20^{+0.06}_{-0.04}$ & $326.1^{+59}_{-37}$ & $30^{+8}_{-5}$ & $9.8^{+0.3}_{-0.2}$ & $0.57$ & $597.5^{+204}_{-133}$ & $62^{+24}_{-15}$ & $6^{+10}_{-4}$ \\
NGC 6101 & $2.90 \pm 0.21$ & $0.96^{+0.30}_{-0.21}$ & $1085.6^{+370}_{-234}$ & $89^{+40}_{-24}$ & $12.0^{+0.6}_{-0.5}$ & $0.47$ & $1660.8^{+852}_{-509}$ & $142^{+81}_{-47}$ & $8^{+18}_{-5}$ \\
NGC 6144 & $2.98 \pm 0.21$ & $0.87^{+0.27}_{-0.19}$ & $1012.2^{+335}_{-213}$ & $84^{+37}_{-23}$ & $11.9^{+0.6}_{-0.5}$ & $0.47$ & $1564.9^{+786}_{-473}$ & $135^{+76}_{-44}$ & $8^{+17}_{-5}$ \\
\begin{tabular}[c]{@{}r@{}}NGC 6171\\ (M107)\end{tabular} & $3.47 \pm 0.22$ & $0.51^{+0.15}_{-0.11}$ & $670.5^{+184}_{-118}$ & $58^{+22}_{-14}$ & $11.1^{+0.5}_{-0.4}$ & $0.51$ & $1102.7^{+487}_{-302}$ & $102^{+50}_{-30}$ & $7^{+14}_{-5}$ \\
\begin{tabular}[c]{@{}r@{}}NGC 6205 \\ (M13)\end{tabular} & $4.19 \pm 0.24$ & $0.23^{+0.07}_{-0.05}$ & $366.8^{+72}_{-46}$ & $34^{+10}_{-6}$ & $10.0^{+0.3}_{-0.3}$ & $0.55$ & $660.3^{+236}_{-153}$ & $67^{+27}_{-17}$ & $7^{+10}_{-4}$ \\
NGC 6362 & $2.95 \pm 0.21$ & $0.91^{+0.28}_{-0.20}$ & $1039.3^{+348}_{-221}$ & $86^{+38}_{-23}$ & $12.0^{+0.6}_{-0.5}$ & $0.47$ & $1600.4^{+811}_{-486}$ & $138^{+78}_{-45}$ & $8^{+17}_{-5}$ \\
NGC 6401 & $3.17 \pm 0.22$ & $0.71^{+0.22}_{-0.16}$ & $865.3^{+268}_{-171}$ & $73^{+30}_{-19}$ & $11.6^{+0.5}_{-0.4}$ & $0.49$ & $1369.6^{+656}_{-399}$ & $121^{+65}_{-38}$ & $8^{+16}_{-5}$ \\
NGC 6426 & $2.55 \pm 0.20$ & $1.41^{+0.45}_{-0.32}$ & $1458.8^{+559}_{-349}$ & $117^{+57}_{-34}$ & $12.7^{+0.7}_{-0.6}$ & $0.44$ & $2135.0^{+1192}_{-696}$ & $174^{+107}_{-61}$ & $9^{+20}_{-6}$ \\
NGC 6496 & $3.47 \pm 0.22$ & $0.51^{+0.15}_{-0.11}$ & $672.9^{+185}_{-119}$ & $58^{+22}_{-14}$ & $11.1^{+0.5}_{-0.4}$ & $0.51$ & $1106.0^{+489}_{-303}$ & $102^{+51}_{-30}$ & $7^{+14}_{-5}$ \\
\begin{tabular}[c]{@{}r@{}}IC 1276 \\ (Pal 7)\end{tabular} & $3.03 \pm 0.21$ & $0.83^{+0.26}_{-0.18}$ & $972.6^{+317}_{-201}$ & $81^{+35}_{-22}$ & $11.8^{+0.6}_{-0.5}$ & $0.48$ & $1512.7^{+751}_{-453}$ & $132^{+73}_{-42}$ & $8^{+17}_{-5}$ \\
NGC 6569 & $4.43 \pm 0.24$ & $0.18^{+0.05}_{-0.04}$ & $299.3^{+51}_{-32}$ & $28^{+7}_{-5}$ & $9.7^{+0.3}_{-0.2}$ & $0.56$ & $555.4^{+183}_{-120}$ & $58^{+22}_{-14}$ & $6^{+9}_{-4}$ \\
NGC 6584 & $3.94 \pm 0.23$ & $0.31^{+0.09}_{-0.07}$ & $451.5^{+101}_{-64}$ & $40^{+13}_{-8}$ & $10.4^{+0.4}_{-0.3}$ & $0.53$ & $787.9^{+304}_{-194}$ & $77^{+34}_{-21}$ & $7^{+11}_{-4}$ \\
\begin{tabular}[c]{@{}r@{}}NGC 6656\\ (M22)\end{tabular} & $4.30 \pm 0.24$ & $0.21^{+0.06}_{-0.04}$ & $335.2^{+62}_{-39}$ & $31^{+9}_{-6}$ & $9.9^{+0.3}_{-0.2}$ & $0.56$ & $611.4^{+211}_{-137}$ & $63^{+25}_{-16}$ & $7^{+10}_{-4}$ \\
NGC 6712 & $3.92 \pm 0.23$ & $0.31^{+0.09}_{-0.07}$ & $459.2^{+103}_{-66}$ & $41^{+13}_{-9}$ & $10.4^{+0.4}_{-0.3}$ & $0.53$ & $799.3^{+310}_{-198}$ & $78^{+34}_{-21}$ & $7^{+12}_{-4}$ \\
NGC 6723 & $3.65 \pm 0.23$ & $0.42^{+0.13}_{-0.09}$ & $577.7^{+147}_{-95}$ & $51^{+18}_{-11}$ & $10.8^{+0.4}_{-0.3}$ & $0.52$ & $971.5^{+409}_{-256}$ & $92^{+43}_{-26}$ & $7^{+13}_{-4}$ \\
\begin{tabular}[c]{@{}r@{}}NGC 6779\\ (M56)\end{tabular} & $3.73 \pm 0.23$ & $0.57^{+0.12}_{-0.09}$ & $543.1^{+134}_{-86}$ & $48^{+17}_{-11}$ & $10.7^{+0.4}_{-0.3}$ & $0.52$ & $921.8^{+380}_{-239}$ & $88^{+41}_{-25}$ & $7^{+13}_{-4}$ \\
\begin{tabular}[c]{@{}r@{}}NGC 6809 \\ (M55)\end{tabular} & $3.33 \pm 0.22$ & $0.60^{+0.18}_{-0.13}$ & $756.1^{+220}_{-141}$ & $64^{+25}_{-16}$ & $11.3^{+0.5}_{-0.4}$ & $0.50$ & $1221.2^{+561}_{-341}$ & $110^{+57}_{-34}$ & $8^{+15}_{-5}$ \\
Pal 11 & $2.65 \pm 0.21$ & $1.26^{+0.40}_{-0.28}$ & $1337.2^{+495}_{-311}$ & $108^{+52}_{-31}$ & $12.5^{+0.7}_{-0.5}$ & $0.45$ & $1982.8^{+1081}_{-635}$ & $164^{+99}_{-56}$ & $9^{+19}_{-6}$ \\
NGC 6934 & $3.70 \pm 0.22$ & $0.40^{+0.12}_{-0.09}$ & $555.6^{+139}_{-89}$ & $49^{+17}_{-11}$ & $10.8^{+0.4}_{-0.3}$ & $0.52$ & $939.8^{+390}_{-245}$ & $89^{+42}_{-25}$ & $7^{+13}_{-4}$ \\
\begin{tabular}[c]{@{}r@{}}NGC 6981\\ (M71)\end{tabular} & $2.98 \pm 0.21$ & $0.87^{+0.27}_{-0.10}$ & $1010.6^{+334}_{-212}$ & $84^{+37}_{-22}$ & $11.9^{+0.6}_{-0.5}$ & $0.47$ & $1562.8^{+785}_{-472}$ & $135^{+75}_{-44}$ & $8^{+17}_{-5}$ \\ \hline
\end{tabular}
\begin{tablenotes}
       \item[a] Is the density of the BHS estimated from the average surface brightness ($\rm L_{\odot} \rm pc^{-2}$) using Equation \ref{fun}.
\item[b] Size of the BHS in pc which is defined as the size in which 50 per cent of the cumulative mass is in BHs while the remaining 50 per cent is in other stars.
\item[c] Is the estimated mass of the BHS system. This correlates with the size of the BHS.
\item[d] Number of BHs in the BHS subsystem.
\item[e] Average mass of BHs in the BHS.
\item[f] Mean mass of stars which are not BHs in the BHS. This anti-correlates with the mean mass of BHs in the BHS. Values were estimated by estimating a linear fit for GC models in which BHs in the BHS had average mass less than 15 $\rm M_{\odot}$.
\item[g] Estimated mass of all BHs in the GC.
\item[h] Predicted number of all BHs in the GC.
\item [i] Potential number of BHs in binary systems with other stars or BHs in these GCs.
       
\end{tablenotes}       
\end{threeparttable}
\end{table*}

\section{Discussion \& Conclusions}\label{last-section}

In Tables \ref{table-obs} and \ref {table-bhs}, we present a list of 
29 Galactic GCs that are likely to be hosting a BHS. For 
Galactic GCs  NGC 4372 and NGC 6101, there have already been 
suggestions that their observed properties could be explained by the 
presence of a BHS. \citet{wang2016} in their million body 
simulation had setup initial conditions for their two models that 
retained about 250 and 1000 BHs at 12 Gyr approximately based on present-day 
observational properties of NGC 4372. \citet{peuten2016} provide 
detailed arguments backed by \textit{N}-body simulations on the 
likelihood that NGC 6101 is hosting a sizeable population of BHs. Our 
work supports their results and through our correlations we find that 
NGC 6101 could be harbouring 100 to 250 BHs with total mass in BHs as high as a few thousand solar masses.

\citet{webb2017} measured how the slope of the mass function (MF) 
changes with respect to the distance from the center of the cluster for five Galactic GCs and compared observed results with those predicted 
from \textit{N}-body simulations. They concluded that both NGC 6101 
and NGC 5466 show little mass segregation and have nearly flat MFs and one possibility to explain this could be the presence of a large number of BHs in these GCs. \cite{ema2015} also found that NGC 6101 shows little evidence for mass segregation. 
In contrast, using mass function slopes obtained from \textit{Hubble Space Telescope}/ACS data \citep{sara2007},
\cite{bs2017} 
suggested that NGC 6101 shows mass segregation that cannot be explained with a BH retention fraction larger than 50 per cent.

Our results show that observable properties of NGC 6101 (V-band magnitude, half-light radius, observable core radius) can be reproduced with a model with just 20 per cent BH retention. For an initially unsegregated \textsc{mocca} GC model with $N=7.0 \times 10^{5}$ initial objects (binary fraction of 10 per cent, mass function given by \citet{kroupa2001}, $W_{0}=3$ (King concentration parameter), half-mass radius of 4.8 pc, tidal radius of 120 pc and Galactocentric radius of 11.7 kpc), 20 per cent BH retention corresponds to about 250 BHs in the GC at 30 Myr. Such a GC has an initial half-mass relaxation time of the order 2.5 Gyr and sustains about 140 BHs till 12 Gyr and can show signs of weak mass segregation. An initially massive GC with large half-mass radius will slowly evolve its BHS compared to an equally massive model with smaller half-mass radius. GCs with a more dynamically evolved, low mass BHS can show stronger signs of mass segregation. This picture can be certainly more complicated if we take into account GCs with different masses, central concentrations and binary fractions. Observed mass segregation might not give us a complete picture of the magnitude of the BHS in a GC unless very deep observations of the GC core are available.  

For NGC 5466, we estimate that there could be an extremely massive BHS of a few thousand solar masses. Based on results from \textit{N}-body simulations, it has been suggested that NGC 288 and NGC 5466 could be containing an IMBH 
\citep{lutz13}. Our estimated properties for the mass of the BHS shows 
that BHS masses in these two clusters could be as large as 3000 
and 5000 $\rm M_{\odot}$. Presence of such massive subsystems in GCs has been suggested in earlier works \citep[see][and reference therein]{arcasedda2016}. We would argue that the CSB for NGC 288 and NGC 5466  are not high enough to be comparable to simulated \textsc{mocca} GC models with massive IMBH \citepalias{aag} therefore it could be likely that these GCs are harbouring a BHS. NGC 5466 and NGC 288 are also particularly interesting as they show presence of extended tidal tails \citep{tails-5466,tails2,lmc2000}. Presence of such tails is typically indicative of strong mass loss suggesting that the GC is filling its tidal radius which could explain the large values for its core and half-light radius. However, it has been suggested that NGC 5466  lost most of its mass due to tides when it crossed the Galactic disc near perigalacticon \citep{fel2007}. While the GC has undergone mass loss, it could still be dynamically young and may contain a BHS. For NGC 288, there is strong evidence that its dynamical evolution has been strongly influenced by the Galactic tidal field. Tidal tails are seen around this GC and there are indications that it may have recently undergone a tidal shock from the disc and the bulge \citep{lmc2000}. Therefore, it may be possible that the observational properties of NGC 288 have been strongly shaped by these tidal shocks rather than the presence of a BHS. NGC 5897 is also a GC which shows tidal tails \citep{balbinot18} and may have undergone tidal interactions with the galactic plane which could also account for its observational properties.

Observed variations of radial velocity measurements of a peculiar star in NGC 3201 which was observed with the MUSE IFU spectrograph has 
provided strong evidence for the presence of a BH in a detached binary system in this GC \citep{giesers2018}. NGC 3201 is one the GCs that meets our selection criterion for Galactic GCs that could potentially be harbouring a large number of BHs. We estimate that there could be as many as 20 BHs in binary systems within this GC and there could be between 80 to 175 single BHs in this GC. Recently \citet{kremer2018b} modelled NGC 3201 using a Monte-Carlo code and showed that its observational properties are in agreement with a GC model containing around 200 or more BHs.

\citet{baumgardt2010} showed evidence for 2 distinct populations of GCs with Galactocentric radius larger than 8 kpc with one group being compact, tidally underfilling while the other group is 
tidally-filling. They argued that the compact clusters were initially 
compact as well whereas most of the tidally filling clusters formed 
with larger half-mass relaxation times. \citetalias{aag} and previous 
works have shown that clusters with larger initial half-mass radii are 
more likely to retain BHs up to 12 Gyr. Comparing the Galactic GC in our shortlist with their results, we find that NGC 288, NGC 5466, IC 4499, 
NGC 6101, NGC 6426 are on their list for tidally-filling clusters. On the other hand, NGC 3201, NGC 4590, NGC 5272, NGC 6205,  
NGC 6779, NGC 6934 and NGC 6981 are on their list for clusters that are compact and could initially have been compact too. If such GCs were initially compact and had large tidal radii then escape velocities at earlier times would be higher which could result in high BH retention (50 per cent). This would mean that  a GC with initially $N=7 \times 10^{5}$ objects could retain about 700 BHs after 20 Myr. We find that in such \textsc{mocca} models the core collapses by 500 Myr forming a BHS with tens of binary BHs that support the evolution of the GC. Gradually the BHS depletes its population of BHs in strong interactions over the course of the GC evolution but such GC models can still have more than a hundred BHs at 12 Gyr. Therefore, even GCs that started out as being relatively compact can still have a sizeable population of BHs at 12 Gyr.

NGC 6656 (M22) harbours two radio detected accreting BH candidates \citep{strader2012} and barely makes it to our shortlist for possible GCs with BHS (its CSB just meets our criterion of being $\lesssim 1 \times 10^{4}$ $\rm L_{\odot}$ $\rm pc^{-2}$). We predict that the mass of the BHS in M22 is about 335 $\rm M_{\odot}$ (corresponding to about 30 BHs) and mass of all BHs should be about 700 $\rm M_{\odot}$. The estimated size of the BHS in M22 is between 0.16 to 0.27 pc. Both the projected position of the BH candidates in M22 are 0.4 and 0.25 pc. As we estimate a low BHS mass for M22 it should have a higher fraction of its BHs in binary systems. Based on our results, the detection of these candidates in a GC with observational properties consistent with GC models that sustain large populations of BHs points towards the fact that there could be many more BHs in M22.

Other GCs within 17 kpc from the Galactic center which did not make it to shortlist but could contain a BHS include:
\begin{itemize}
 \item NGC 6121, NGC 6218, NGC 6254 (M10), NGC 6287, Palomar 6, Djorg 1, Djorg 2, HP1 and NGC 6749  could possibly contain BHS of a few hundred solar masses. These GCs were not shortlisted as present-day half-mass relaxation times were lower than 0.9 Gyr or the value was not available for these GCs. Recently, \citet{Shishkovsky2018} reported the detection of a red straggler binary with an invisible companion in NGC 6254 (M10). \citet{Shishkovsky2018} also found an X-ray counterpart to this binary and opined that the  invisible companion in this binary could be a BH. From our correlations we predict that M10 could contain  $75^{+32}_{-20}$ BHs with up to 19 BHs being in binary systems.
\item Palomar 12, Terzan 1, BH 176, Terzan 3, Terzan 7, Terzan 12, NGC 6366, NGC 6558, NGC 6235, NGC 2298 and Palomar 6 did not make it to the shortlist because they have magnitude values larger than -6.5 and/or present day half-mass relaxation times were lower than 0.9 Gyr. Palomar 12, Terzan 1, BH 176, Terzan 7, Terzan 12 are too dim compared to most BHS models. These GCs could be dark star clusters with a BHS \citep{ban2011}. On the other hand, GCs NGC 6366, NGC 6558, NGC 6235, NGC 2298 and Palomar 6 are sufficiently bright and could contain a massive BHS.
\item There are a few GCs that had CSB values slightly larger than our maximum cutoff value of of $\sim 1 \times 10^{4}$ $\rm L_{\odot}\rm pc^{-2}$ that could also host a BHS containing up to a few hundred solar masses. These include NGC 6287, NGC 6380, NGC 6356, NGC 6333 (M9) and NGC 6553, NGC 6637 (M69) and NGC 6760.
\item There are a few more additional GCs for which the average surface brightness inside the half-light radius agrees with models with BHS but did not make the shortlist because of CSB values being much higher than our maximum cutoff value. There are NGC 5904, NGC 5946, NGC 6256, NGC 6626, NGC 6539, NGC 6284, NGC 6293, NGC 5927, NGC 6304, NGC 6522, NGC 6752, NGC 6642, NGC 6355, NGC 6341 (M92), NGC 6864 (M75) and NGC 6139. The correlations that we used found a substantial BHS in these GCs, with masses of a few hundred solar mass. However, only a few simulated GC models having a large number of BHs at 12 Gyr have CSB values which are as high as CSB values for these GCs.
\item Both NGC 5139 ($\rm \omega$ Cen) and NGC 6402 (M14) have low CSB values but brightness (total V-band magnitude) larger than any simulated GC model at 12 Gyr and hence did not make it to the shortlist. Our correlations do not predict massive BHS for these GCs. For $\rm \omega$ Cen, we estimate total number of BHs to be between 50 and 80. 
\end{itemize}

GCs with distances larger than 17 kpc from the Galactic center which could be hosting a sizeable number of BHs include:

\begin{itemize}
 \item NGC 5053 and NGC 7492.
 \item Applying our correlations to very distant Galactic GCs like Palomar 14, AM 1, Palomar 3, Palomar 15, Eridanus and Pal 4 also yields substantial BHS mass. Most of these GCs are not bright (absolute magnitdus are around -4).  As discussed in Section \ref{sec2:method}, we do not consider these GCs as none of the simulated GC models have such large Galactocentric radii. However, this does not rule out the possibility that GCs at large distances from the center of the Galaxy do not harbour BHS.
\end{itemize}

Our results show that about 20 per cent of Galactic GCs could potentially be hiding many BHs. The results from this study can be useful for observers trying to identify elusive BH candidates in GCs. The estimated number of BHs in binaries for the GCs we have identified in this paper is rather low (from few tens to several dozens at best). The likelihood of the BHs in these binaries to be accreting is even lower which explains the difficulty in detecting such systems. The large presence of single BHs estimated for a few of these shortlisted GCs could possibly be detected via microlensing \citep{Paczynski94,bennet2002,pt2012,minn2015,lu16,wyrz2016,rybicki2018}.

In addition to the limitations and cautions specified in Section \ref{limits}, it is important to point out that the correlations from \citetalias{aag} come from GC models simulated with the \textsc{mocca} code. Similar to other numerical simulation codes (e.g., \textit{N}-body or other Monte Carlo codes) that compute the evolution of star clusters, there are many limitations and important physical processes that are not taken into account which could be important for the evolution of a real Galactic GC. These include rotation, proper Galactic potential and tides, GC's orbit and its formation environment. Therefore, the results presented in Table \ref{table-bhs} are only estimates and we would like to stress that our results do not imply that Galactic GCs other than the ones identified in this study do not harbour stellar mass BHs. The possibility of several BHs existing in other Galactic GCs cannot be ruled out. The 29 Galactic GCs identified in this study are likely to be harbouring a significantly large number of BHs and have observational properties similar to simulated GC models that sustain large number of BHs. In the future we plan on using a similar approach to be able to identify Galactic GCs with IMBHs.

\section*{Acknowledgements}
We would like to thank the anonymous reviewer for the comments and suggestions they provided that helped in improving this manuscript. AA and MG acknowledge support from National Science Center (NCN), 
Poland, through the grant UMO-2016/23/B/ST9/02732. AA was also 
supported by NCN, Poland through the grant UMO-2015/17/N/ST9/02573. 
MAS acknowledges the Sonderforschungsbereich SFB 881 "The Milky Way 
System" (subproject Z2) of the German Research Foundation (DFG) for 
the financial support. 

%%%%%%%%%%%%%%%%%%%%%%%%%%%%%%%%%%%%%%%%%%%%%%%%%%



%\bibliographystyle{mnras}


%%%%%%%%%%%%%%%%%%%%%%%%%%%%%%%%%%%%%%%%%%%%%%%%%%

%%%%%%%%%%%%%%%%% APPENDICES %%%%%%%%%%%%%%%%%%%%%

\appendix

\section{Table with Correlations}\label{app}

Table \ref{tab:app-table} shows all the correlations and fitted parameters that were used to obtain the estimates for BHS properties and overall populations of BHs in 29 GCs in Section \ref{result}. Some of the correlations are slightly different from the ones presented in \citetalias{aag} (see Table \ref{tab:app-table} notes for details), for those correlations improved fittings linear in log-log were obtained by limiting the range of the dependent parameter to parameters estimated for the 29 shortlisted GCs.

\begin{table*}
\renewcommand{\arraystretch}{1.5}
\begin{threeparttable}
\caption{In this table we define all the correlations that were used to obtain the values provided in Table \ref{table-bhs}. By knowing the V-band luminosity and half-light of a given GC that meets the criteria mentioned in Section \ref{criterion}, we can estimate the properties of its BHS using these correlations.}
%\begin{adjustbox}{max width=\textwidth}
	
	\label{tab:app-table}
    
\begin{tabular}{ccl}
\hline
\textbf{\begin{tabular}[c]{@{}c@{}}Correlated \\ Parameters\end{tabular}} & \textbf{\begin{tabular}[c]{@{}c@{}}Correlation\\  Used\end{tabular}} & \multicolumn{1}{c}{\textbf{\begin{tabular}[c]{@{}c@{}}Fitted\\  Parameters\end{tabular}}} \\ \hline
$[L_{\rm V}/r_{hl}^{2}]$ ---  $\rho^{}_{\bhs}$ \tnote{a} & \multicolumn{1}{c|}{Log{[}$\rho^{}_{\bhs}${]} = $A$ Log$[L_{\rm V}/r_{hl}^{2}]$ + $B$} & \multicolumn{1}{l|}{\begin{tabular}[c]{@{}l@{}}A = $1.43 \pm 0.03$\\ B = $-1.91 \pm 0.11$\end{tabular}} \\ \hline
$\rho^{}_{\bhs}$ --- $R_{\bhs}$ \tnote{b} & \multicolumn{1}{c|}{Log{[}$R_{\bhs}${]} = (Log$[\rho^{}_{\bhs}]$ $-$ D) / C} & \multicolumn{1}{l|}{\begin{tabular}[c]{@{}l@{}}C = $-2.11 \pm 0.07$\\ D = $2.86 \pm 0.03$\end{tabular}} \\ \hline
$R_{\bhs}$ --- $M_{\bhs}$ \tnote{c} & \multicolumn{1}{c|}{Log{[}$M_{\bhs}${]} =E Log{[}$R_{\bhs}${]} + F} & \multicolumn{1}{l|}{\begin{tabular}[c]{@{}l@{}}E = $0.77 \pm 0.07  $\\ F = $3.05 \pm 0.03$\end{tabular}} \\ \hline
$M_{\bhs}$ --- $N_{\bh}$ in BHS ($N_{\bhs}$) \tnote{d} & \multicolumn{1}{c|}{Log{[}$N_{\bhs}${]} = G Log{[}$M_{\bhs}${]} + H} & \multicolumn{1}{l|}{\begin{tabular}[c]{@{}l@{}}G = $0.903 \pm 0.008 $\\ H = $-0.79 \pm 0.02 $\end{tabular}} \\ \hline
Mean $M_{\bh}$ in BHS ($m_{\bhs}$) --- $R_{\bhs}$ \tnote{e} & \multicolumn{1}{c|}{Log{[}$m_{\bhs}${]} = I Log{[}$R_{\bhs}${]} + J} & \multicolumn{1}{l|}{\begin{tabular}[c]{@{}l@{}}I = $0.13 \pm 0.01$\\ J = $1.093 \pm 0.005$\end{tabular}} \\ \hline
Mean $M_{\bh}$ in BHS ($m_{\bhs}$) ---Mean $M_{\star}$ in BHS ($m_{\star}$) \tnote{f} & \multicolumn{1}{c|}{$m_{\star}$ = K $m_{\bhs}$  + L} & \multicolumn{1}{l|}{\begin{tabular}[c]{@{}l@{}}K = $-0.040 \pm 0.003$\\ L = $0.95 \pm 0.04$\end{tabular}} \\ \hline
$M_{\bh}$ in GC ($M_{\rm ALL-BH}$) --- $M_{\bhs}$ \tnote{g} & \multicolumn{1}{c|}{Log{[}$M_{\rm ALL-BH}${]} = M Log{[}$M_{\bhs}${]} + N} & \multicolumn{1}{l|}{\begin{tabular}[c]{@{}l@{}}M = $0.85 \pm 0.01$\\ N = $0.64 \pm 0.04$\end{tabular}} \\ \hline
$N_{\bh}$ in GC ($N_{\rm ALL-BH}$) --- $N_{\bhs}$ \tnote{h} & \multicolumn{1}{c|}{Log{[}$N_{\rm ALL-BH}${]} = O Log{[}$N_{\bhs}${]} + P} & \multicolumn{1}{l|}{\begin{tabular}[c]{@{}l@{}}O = $0.77 \pm 0.02$\\ P = $0.65 \pm 0.65$\end{tabular}} \\ \hline
$N_{\bh}$ in Binaries ($N_{\rm BHB}$) ---$M_{\bhs}$ \tnote{i} & \multicolumn{1}{c|}{Log{[}$N_{\rm BHB}$/$N_{\rm ALL-BH}${]} = Q Log{[}$M_{\bhs}${]} + R} & \multicolumn{1}{l|}{\begin{tabular}[c]{@{}l@{}}Q= $-0.50 \pm 0.04$\\ R = $0.28 \pm 0.11$\end{tabular}} \\ \hline
\end{tabular}
\begin{tablenotes}
\item[a] Gives the correlation between the observed V-band luminosity, half-light radius and the density of the BHS. $\rm L_{\rm V}$ is in units of solar luminosity, $\rm r_{hl}$ is in the units of pc. The estimated BHS mass density is in units $\rm M_{\odot}$ $\rm pc^{-3}$.
\item[b] Shows the anti-correlation between the density of the BHS estimated in the previous correlation and the size of the BHS which is units of pc.
\item[c] Relation between the size of the BHS and the mass of the BHs in BHS in units of solar mass.
\item{d} Relation between the total mass of the BHs in the BHS and the number of BHs in the BHS.
\item[e] Relation between the average mass of BHs in the subsystem and the size of the BHS in units of pc.
\item[f] Anti-correlation between the average of stars in the BHS and the average of mass of BHs in the BHS. This relation is different from the one shown in \citepalias{aag} and was found by limiting to models which had average BH mass in BHS lower than 14 $\rm M_{\odot}$ (as was the case for the 29 shortlisted Galactic GCs).
\item[g] Correlation between the total mass of all BHs in the GC and the mass of the BHs in the BHS in units of solar mass.
\item[h] Relation for the total number of all BHs in the GC and the number of BHs in the BHS. 
\item[i] Shows the anti-correlation between the fraction of BHs in  binary systems in the GC and the mass of BHs in the BHS in units of solar mass. $N_{\rm BHB}$ includes all binaries in which at least one component is a BH. This relation was also obtained by limiting ourselves to BHS masses in the range of 400 to 3000 $\rm M_{\odot}$. The full relation can be found in \citetalias{aag}.
\end{tablenotes}       
\end{threeparttable}
%\end{adjustbox}
\end{table*}

%%%%%%%%%%%%%%%%%%%%%%%%%%%%%%%%%%%%%%%%%%%%%%%%%%

% Don't change these lines
\bsp	% typesetting comment
\label{lastpage}
\end{document}